\documentclass[aps,prb,reprint,longbibliography,superscriptaddress]{revtex4-2}

\usepackage{graphicx}   
\usepackage{setspace}
\usepackage{dcolumn}    
\usepackage{bm}         
\usepackage{amsmath}    
\usepackage{amssymb}    
\usepackage{mathtools}
\usepackage{amsfonts}
\usepackage{xcolor}
\usepackage{algorithmicx}
\usepackage{algpseudocode}

\newcommand{\abs}[1]{\left\vert#1\right\vert}
\newcommand{\ket}[1]{{\left| #1 \right>}}

\newcommand{\expec}[3]{{\left< #1 \middle\vert #2 \middle\vert #3 \right>}}
\newcommand{\bracket}[1]{{\left< #1 \right>}}

\begin{document}

\title{Imaginary-time correlations in time-sliced stochastic series expansion}

\author{Ryan Flynn}
\thanks{These authors contributed equally to this work.}
\affiliation{Department of Physics, Boston University, 590 Commonwealth Avenue, Boston, Massachusetts 02215, USA}
\email{rflynn22@bu.edu}

\author{Adam Iaizzi}
\thanks{These authors contributed equally to this work.}
\affiliation{Department of Physics and Center for Theoretical Physics, National Taiwan University, Taipei 10607, Taiwan}

\author{Sibin Yang}
\affiliation{School of Physical and Mathematical Sciences, Nanyang Technological University, 21 Nanyang Link, Singapore 637371}

\author{Ying-Jer Kao}
\affiliation{Department of Physics and Center for Theoretical Physics, National Taiwan University, Taipei 10607, Taiwan}

\author{Anders W. Sandvik}
\affiliation{Department of Physics, Boston University, 590 Commonwealth Avenue, Boston, Massachusetts 02215, USA}

\date{\today}

\begin{abstract}
Combined with numerical analytic continuation techniques, quantum Monte Carlo (QMC) methods enable the extraction of real-frequency dynamical properties from imaginary-time correlation functions. However, the efficient computation of imaginary-time correlation functions by QMC simulations can (depending on the particular model used) be challenging, particularly for operators that are off-diagonal in the computational basis. In this work, we present an efficient and general algorithm within the stochastic series expansion (SSE) framework for evaluating imaginary-time correlation functions of both diagonal and off-diagonal operators. The algorithm builds on a discrete imaginary-time slicing of the SSE operator string, which provides correlation functions on a grid of well-defined imaginary-time points with no discretization error. For off-diagonal operators, we derive estimators that integrate directly into the existing SSE directed-loop or cluster updating schemes, introducing only minimal computational overhead. We benchmark the method on the one-dimensional transverse-field Ising model (sampling with cluster updates) and XXZ spin chain (using directed-loop sampling), demonstrating excellent agreement (with only statistical errors) with exact diagonalization of small systems. We also study larger systems to demonstrate efficiency.
\end{abstract}

\maketitle
\section{Introduction}\label{Sec:Intro}

Quantum Monte Carlo (QMC) methods are among the most accurate and unbiased numerical tools for studying strongly correlated quantum many-body systems \cite{SandvikReview,GubernatisBook}. In this work, we consider the evaluation of imaginary-time dependent correlation functions with the stochastic series expansion (SSE) \cite{Sandvik_PRB_1991,Sandvik_JPA_1992,Sandvik_PRB_1997,SandvikReview}, a widely used QMC method for sign-problem-free \cite{Loh_PRB_1990,Henelius_PRB_2000,Troyer_PRL_2005} Hamiltonians that samples a power-series representation of the partition function and provides a natural discrete but exact imaginary-time representation of quantum statistical mechanics at arbitrary temperatures, including the limit $T\to 0$ at finite size. SSE admits efficient loop \cite{Sandvik_PRB_1999,Syljuasen_PRE_2002,Syljuasen_PRE_2003,Melko_PRL_2008,Desai_PRB_2021,Takahashi_2024a,Dao_2026} and cluster \cite{Sandvik_PRE_2003,Sandvik_PRL_2002,Melko_PRE_2005,Biswas_PRB_2016,Yan_PRB_2019,Patil_SCI_2026} updates, generalizing classical global importance sampling schemes \cite{Swendsen_PRL_1987,Luijten_INT_1995,Evertz_PRL_1993,Evertz_ADV_2003}. The SSE method has been applied to a broad range of models, including (for mainly recent representative examples; see further references therein) bipartite quantum spin systems with competing interactions \cite{Sandvik_PRL_2007,Lou_PRB_2009,Kaul_PRB_2015,Takahashi_PRR_2020,Takahashi_2024b,Kundu_PRB_2024}, bosonic lattice models \cite{Sengupta_PRL_2005,Schaffer_PRB_2009,Pippan_PRA_2009,Wang_PRL_2015,Merali_SCI_2024}, including dynamic phonons \cite{Weber_PRB_2021,Gotz_2024}, as well as some fermion models \cite{Wang_PRB_2016,Li_ANN_2019} and frustrated quantum spin systems \cite{Alet_PRL_2016,Wessel_PRB_2018} for which the sign problem can be avoided.

An important application of QMC is the computation of real-frequency dynamical properties, particularly spectral functions that can be directly related to time-domain experiments such as inelastic neutron scattering and angle-resolved photoemission spectroscopy \cite{Hong_Nature_2017,Jain_Nature_2017}. In addition to direct comparisons with experiments, an important role of QMC calculations is to elucidate the connection between the dynamics of lattice models and related field theories, for which calculations often require approximations that need to be tested \cite{Gazit_PRB_2013,WitczakKrempa_NatPhys_2014,Katz_PRB_2014}. Because QMC simulations operate in imaginary time, direct comparisons of experimental or analytically obtained real-frequency spectral functions require numerical analytic continuation of correlation functions \cite{Jarrell_PR_1996} (though analytical results can also be continued to imaginary time \cite{Katz_PRB_2014}, but many details are lost when only comparing in imaginary time). Modern analytic continuation techniques, including maximum-entropy methods \cite{Silver_PRB_1990,Jarrell_PR_1996,Jarrell2012MaxEnt} and stochastic analytic continuation (SAC) \cite{Sandvik_PRB_1998,BeachSAC,Sandvik_PRE_2016,ShaoSAC}, have made such comparisons increasingly reliable; for example Refs.~\cite{Lohofer_PRB_2015,Shao_PRX_2017,Qin_PRB_2017,Ma_PRB_2018,Yang_2025a}. With these advances in analytic continuation, a remaining bottleneck is the computation of imaginary-time correlation functions themselves, which in general can still be challenging to obtain efficiently in QMC to sufficient numerical precision for reliable analytic continuation. 

Correlation functions of operators diagonal in the computational basis can be accessed relatively easily with SSE QMC \cite{Sandvik_JPA_1992,Sandvik2019SSE} and some important off-diagonal ones can be computed using generalizations of the worm algorithm \cite{Prokofev_JETPL_1996} approach of an enlarged configuration space to SSE \cite{Dorneich_PRE_2001}.  However, in conventional SSE formulations, establishing the mapping from operator string position to imaginary time has typically required a computationally expensive binomial expansion \cite{Sandvik_JPA_1992,Dorneich_PRE_2001} or stochastic sampling of the distribution \cite{Pippan_PRA_2009}. As a result, SSE-based studies of dynamical properties have often been restricted to diagonal structure factors, e.g. $S^{zz}(q,\omega)$, or to models with SU($2$) symmetry where transverse correlators are equal to diagonal ones. Refs.~\cite{Laflorencie_PRB_2004,Syljuasen_PRB_2008} and ~\cite{Pippan_PRA_2009,Grossj_PRB_2009,Rahnavard_PRB_2015} are examples of SSE calculations of off-diagonal static and dynamic correlations, respectively beyond the demonstration \cite{Dorneich_PRE_2001} on the method. 

In this work, we present efficient and general algorithms to compute imaginary-time correlation functions within the SSE framework. Our approach builds on a discrete imaginary-time slicing of the SSE operator string, first implemented in projector QMC \cite{Shao_PRB_2015} and extended to finite-temperature SSE \cite{Sandvik2019SSE}. This yields correlation functions on a grid of well-defined imaginary-time points but introduces no discretization errors. Within this framework, we address three classes of correlation functions. Diagonal correlators can be evaluated directly from propagated states at time-slice boundaries, as previously established \cite{Sandvik2019SSE}. For off-diagonal operators that appear explicitly in the Hamiltonian, we derive a simple operator-counting estimator, requiring no summation over the full operator string. For off-diagonal operators not contained in the Hamiltonian, we construct an estimator by leveraging directed-loop updates \cite{Dorneich_PRE_2001,Syljuasen_PRE_2002} to sample an extended configuration space, in a construction analogous to the continuous-time worm algorithms \cite{Prokofev_JETPL_1996,Prokofev_JETP_1998,Prokofev_PRA_1998,Mishchenko_PRB_2001}. In both off-diagonal cases, the algorithms introduce only minimal computational overhead beyond the standard SSE sampling for equal-time observables.

We benchmark the time-sliced SSE method using the one-dimensional transverse-field Ising model (TFIM), which tests the estimator for off-diagonal operators contained in the Hamiltonian, and the XXZ model with an external magnetic field, which tests both diagonal and off-diagonal correlators of operators not present in the Hamiltonian. In both cases, the algorithm integrates seamlessly into existing SSE implementations with minimal modifications to standard directed-loop updates and measurements. The procedures for these two models are also rather generic and can be easily adapted to a wider range of extended TFIM and XXZ models, their generalizations as well as other models with sufficiently similar operator structure. We note that time-sliced SSE followed by analytic continuation was recently used to compute the single-particle spectral function in quantum magnets with different types of ground states \cite{Yang_2025a,Yang_2025b}, using a special basis including a single hole in addition to the $S=1/2$ spins. The procedures there are particular to the single-hole spectral function and very different from the cases considered here.

The remainder of the paper is organized as follows. In Sec. \ref{Methods} we briefly review the SSE framework and notation. We also introduce the imaginary-time slicing procedure. In Sec.~\ref{DCorr} we review the measurement of diagonal correlation functions in SSE, which serves as the foundation for the off-diagonal algorithms that follow. In Sec.~\ref{OinH} and Sec.~\ref{OnotinH} we derive the estimator for correlation functions of off-diagonal operators that are contained in the Hamiltonian ($\mathcal{O}\in H$) and those that are not ($\mathcal{O}\notin H$) respectively. In Sec.~\ref{Results} we present QMC results for the TFIM and XXZ spin chains. We test agreement with exact diagonalization of small systems for correlators of the different types of operators and also show some results for larger systems to demonstrate efficiency on large scales in space and time. Finally, we conclude in Sec.~\ref{Discussion} with a discussion of prospects for applying this method to outstanding problems in the field. 

\section{Stochastic Series Expansion}\label{Methods}

In this section we first introduce the two models for which we specifically construct algorithms. We then briefly summarize the SSE framework for QMC simulations, following mainly Refs.~\cite{Sandvik_PRB_1999,Syljuasen_PRE_2002,Dorneich_PRE_2001}. Our purpose is not to provide a full review---comprehensive expositions already exist \cite{SandvikReview,GubernatisBook}---but instead to introduce the notation and algorithmic pieces that will be modified in the imaginary-time slicing procedure and in the measurement of correlation functions.

\subsection{Models}\label{sec:models}

We define the standard $S = 1/2$ ferromagnetic 1D TFIM using the Pauli matrices,
\begin{equation}\label{Eq:TFIM}
  H = -\sum_{i}\sigma^z_i\sigma^z_{i+1} - h\sum_i\sigma^x_i 
\end{equation}
where we have set the Ising interaction to $J=1$ and $h$ is the transverse field strength. This model has a self-dual quantum critical point at $h=1$ which separates a trivial paramagnet with only $x$ magnetization for $h>1$ and two-fold degenerate $z$ ferromagnetic state for $h<1$. The transition can be detected in the $\sigma^z_i$ correlations, which are diagonal in the computational basis. The main reason for studying the TFIM here is that it provides a simple example of off-diagonal correlations of operators that appear in the Hamiltonian, i.e., the field terms $\sigma^x_i$. 

The second model is the XXZ Heisenberg chain in an external field, which we define using the spin-$1/2$ operators $S_i^{x,y,z} = \sigma_i^{x,y,z}/2$,
\begin{eqnarray}\label{Eq:XXZ}
  H = && J\sum_i\left[S_i^xS_{i+1}^x + S_i^yS_{i+1}^y + \Delta S_i^zS_{i+1}^z\right] \nonumber \\
      && - h\sum_i S_i^z,
\end{eqnarray}
again with the z components diagonal. For $-1<\Delta\leq1$ the ground state is a gapless Luttinger liquid while $|\Delta|>1$ leads to ferromagnetic or  antiferromagnetic order depending on the sign of $J\Delta$. The point $\Delta = 1$, $h=0$ recovers the isotropic Heisenberg model, where the SU(2) symmetry exactly relates the two operator types. In this case, the operators $S_i^{x,y}$ or $S_i^\pm$ that define the transverse correlations functions do not appear individually in the Hamiltonian and the way these correlations are evaluated is therefore very different from the TFIM case.

Together the two models cover three classes of space-time correlation functions that will be discussed in Secs. \ref{DCorr}-\ref{OnotinH}: diagonal correlators, and off-diagonal correlators of operators contained/not-contained in $H$. SSE simulations of these models also utilize different updating schemes of the quantum fluctuations---generalizations of the Swendsen-Wang-type clusters for the TFIM \cite{Sandvik_PRE_2003} and directed loops \cite{Sandvik_PRB_1999,Syljuasen_PRE_2002} for the XXZ chain---and so together provide a comprehensive test of the algorithms. The two sampling methods and evaluations of correlation functions can also be generalized to a broad range of extended TFIM and XXZ models, as well as other models with analogous operator structure.

\subsection{Series expansion and operator-string representation}

The SSE method is based on a power-series expansion of the partition function,
\begin{equation}\label{Eq:SeriesExp}
    Z = \mathrm{Tr}\left[ e^{-\beta H} \right]
      = \sum_{n=0}^{\infty} \frac{\beta^n}{n!}
        \sum_{\{\alpha\}} \expec{\alpha}{(-H)^n}{\alpha}
\end{equation}
where $\{\alpha\}$ is a chosen computational basis. The Hamiltonian is decomposed into a sum of local operators $H_b$,
\begin{equation}\label{Eq:BondOps}
    H = -\sum_{b} H_b,
\end{equation}
such that acting on a basis state produces only a single other basis state $\ket{\alpha'} = H_b\ket{\alpha}$ (i.e., no branching to a superposition, which can always be accomplished more generally by splitting operators appropriately). In the XXZ model these are all two-body terms, but in the TFIM the transverse field contributes single-site terms. In principle, three-body and higher-order interactions can also be studied, and in some cases it is advantageous to define the local operators on cells larger than the minimum size \cite{Louis_PRB_2004,Sandvik_PRL_2002,Melko_PRE_2005,Biswas_PRB_2016}. If the condition $\expec{\alpha'}{H_b}{\alpha}>0$ can be satisfied, the negative signs in Eqs.~\eqref{Eq:SeriesExp} and \eqref{Eq:BondOps} cancel and importance sampling without a sign problem is possible \cite{Henelius_PRB_2000,Troyer_PRL_2005}. In many cases where such a transformation to a positive-definite sampling space is in principle possible, it can also be implemented implicitly for the off-diagonal operators when the contributing operator strings must necessarily contain an even number of these operators with $\expec{\alpha'}{H_b}{\alpha}<0$, e.g., in the case of the XXZ model on bipartite lattices. The signs of the diagonal matrix elements can always be made negative (so that the weights are positive) by subtracting a suitable constant. 

Expanding $(-H)^n$ produces an ordered sequence of local operators, or the \emph{operator string}, 
\begin{equation}
    \mathcal{S}_n = (H_{b_1}, H_{b_2}, \ldots, H_{b_n}),
\end{equation}
with $H_{b_p}$ drawn from the set in Eq.~\eqref{Eq:BondOps}. This operator string acts on one of the basis states $\ket{\alpha}$ in Eq.~\eqref{Eq:SeriesExp} and the unique propagation (i.e., with no branching into a superposition of states) must lead back to the same initial state. In practice, the series expansion is often truncated at a cutoff $M$ sufficiently large such that $n$ never approaches $M$ and, therefore, introduces no approximation. The truncations allow for a constant-length operator string, which leads to simpler detailed-balance conditions and is more efficient computationally. The remaining $M-n$ positions in the string are then filled with randomly distributed identity operators. An SSE configuration is specified by $\mathcal{C} = \{\alpha, S_M\}$, consisting of the initial state $\ket{\alpha}$ and an operator string of fixed length $M$. By a few simple steps, the partition function Eq.~(\ref{Eq:SeriesExp}) can then be written as
\begin{equation}\label{Eq:SSEZ}
    Z = \sum_{\mathcal{C}}\frac{\beta^n (M-n)!}{M!}\expec{\alpha}{\prod_{p=1}^M H_{b_p}}{\alpha}
\end{equation}
where now the operators $H_{b_p}$ are drawn from the set in Eq.~\eqref{Eq:BondOps} amended by the unit operator $\mathbb{I}$. The expansion order $n$ is then implicit, equal to the number of non-unit operators in the string.

\subsection{Diagonal update}

Sampling the expansion order $n$ is achieved through the diagonal update, where diagonal local operators $H_b$ are inserted or removed in the operator string. At position $p\in \{1,\ldots,M\}$, let $\ket{\alpha(p)}$ be the state propagated by the current operator string from the initial configuration $\ket{\alpha}$ through the first $p-1$ operators and consider the cases for which $[H_b] = \expec{\alpha(p)}{H_b}{\alpha(p)}$ is a diagonal matrix element. The Metropolis probabilities to insert a diagonal local operator $H_b$ at a position containing an identity, or to remove a diagonal $H_b$ from the string, are
\begin{subequations}
\label{eq:probs}
\begin{align}
    P(\mathbb{I} \rightarrow H_b) &= \frac{\beta N_b [H_b]}{M - n}, \label{eq:insert} \\
    P(H_b \rightarrow \mathbb{I}) &= \frac{M-n+1}{\beta N_b [H_b]}.
    \label{eq:remove}
\end{align}
\end{subequations}
respectively, where $N_b$ is the total number of lattice units on which the Hamiltonian operators can be placed. If the chosen operator has vanishing matrix element the update is simply rejected. These expressions enforce detailed balance and sample the proper distribution of expansion orders \cite{Sandvik_PRB_1999}.

\subsection{Directed-loop and cluster updates}

\begin{figure}[t] 
  \centering
  \includegraphics[width=\columnwidth]{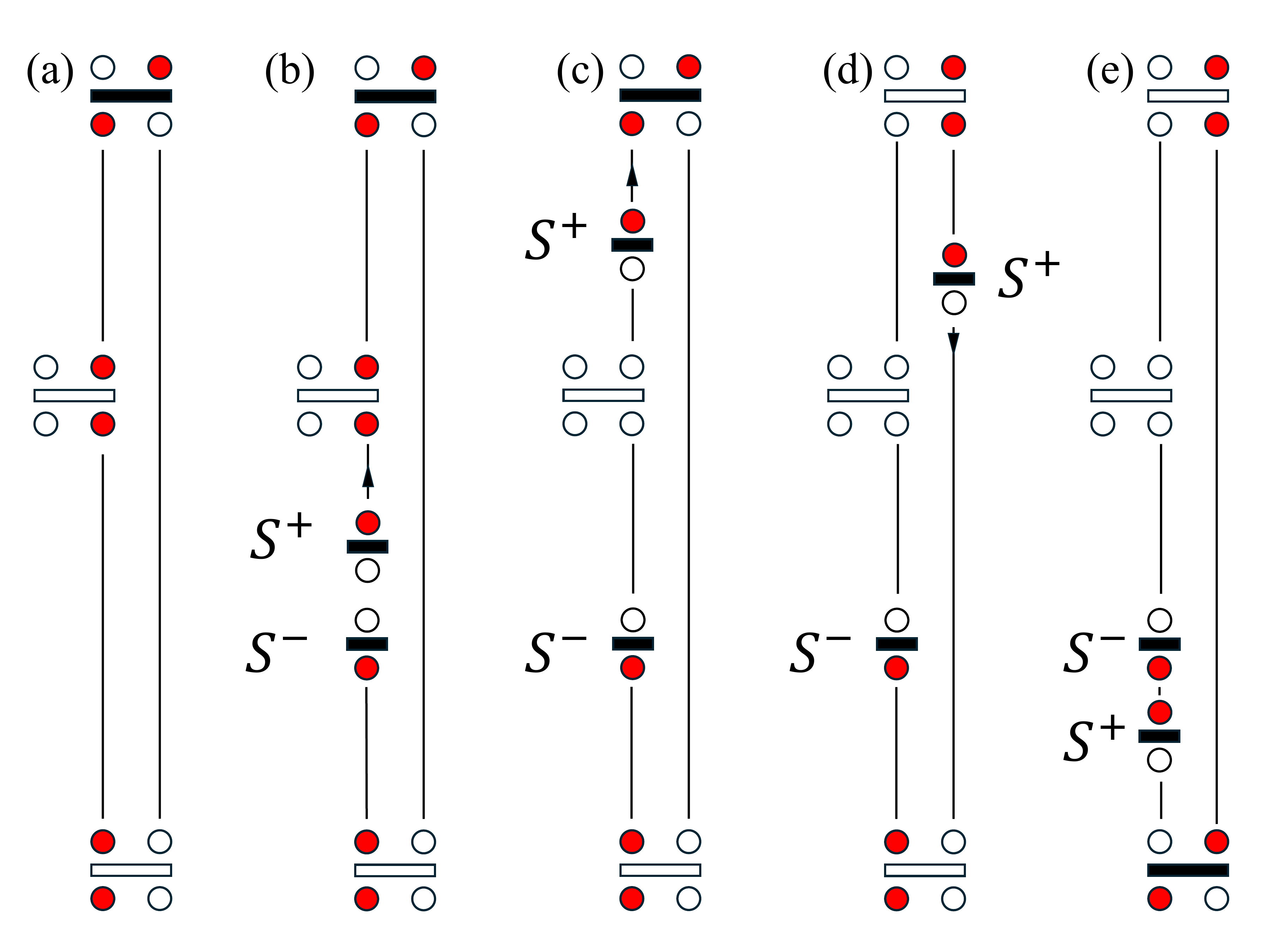}
  \caption{The directed-loop update for the XXZ model can be thought of as the insertion of an $S^+,S^-$ pair of operators at some spacetime point $(r_0,\tau_0)$, and then propagating one of the corresponding defects (the ``head'', here $S^+$) through the configuration. Partial SSE configurations are here shown in the linked-vertex representation \cite{SandvikReview}, where spin states between operators are represented by vertical lines; realized as bidirectional links used in coding. (a) A configuration in the space $\mathcal{C}$ without defects, with \textcolor{red}{$\bullet$}/$\circ$ representing up/down spins, and (black) white rectangles representing (off)-diagonal operators. (b) Construction of a loop is commenced  by inserting an $S^+S^-$ pair on a chosen link. (c)/(d) Propagation of the head along a path according to the directed-loop update, flipping spins and operators. These ``open loop configurations'' belong to the two-defect space $\in\mathcal{C}'$ contributing to the off-diagonal spin correlation functions. (e) Closing of the loop when $S^+$ meets $S^-$ again, thus annihilating the defects and leading back to the configuration space $\in\mathcal{C}$.}
\label{fig:LoopDefect}
\end{figure}

\begin{figure}[t] 
  \centering
  \includegraphics[width=\columnwidth]{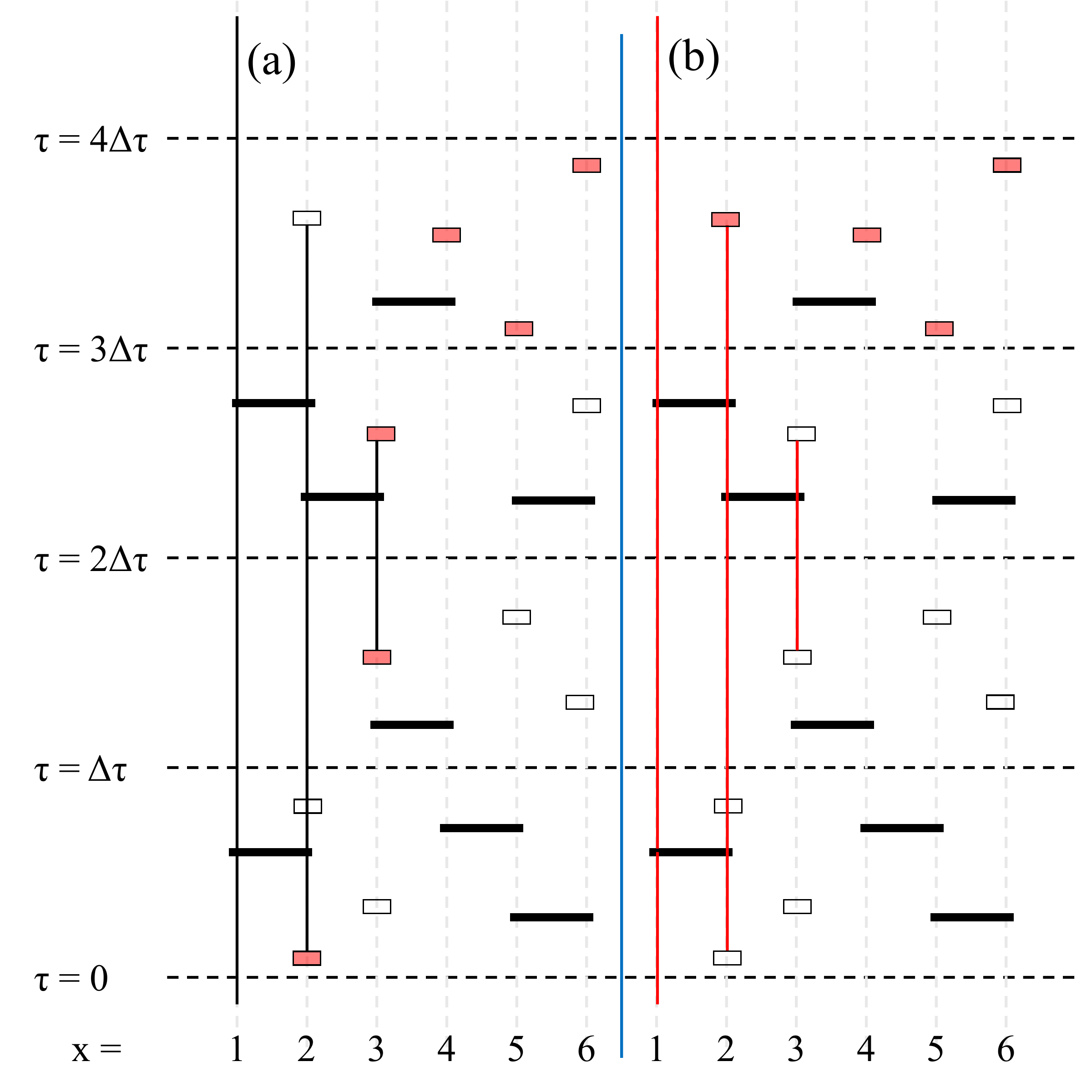}  
  \caption{Illustration of the TFIM cluster update. Ising operators are represented by black rectangles. White/red rectangles represent $h\mathbb{I}$ and $h\sigma^x$  operators respectively, where the identity operators are present for efficient cluster updates. A single cluster (a) is shown, and the action of flipping the cluster (b) is illustrated. Clusters which touch single-site operators flip $h\mathbb{I} \leftrightarrow h\sigma^x$ as shown. }
  \label{fig:Clusterupdate}
\end{figure}

Ergodicity and efficient sampling of off-diagonal operators are achieved through non-local cluster-type updates. For our two benchmark models, the relevant updates are the directed-loop algorithm \cite{Syljuasen_PRE_2002} for the XXZ chain and a Swendsen-Wang-type cluster update \cite{Sandvik_PRE_2003} for the TFIM. We will not discuss the details of these updates here, but some procedures will be closely related to the evaluation of off-diagonal correlations that will be described in the later sections. In both updates, the operator string is viewed as a space--imaginary-time lattice of vertices, and a loop/cluster is constructed that propagates through connected legs of the vertices while flipping local spin states according to detailed balance. Flipping the loop/cluster yields a new valid SSE configuration. An example of a simple directed-loop update of an XXZ configuration is shown in Fig.~\ref{fig:LoopDefect} (here without reference to the time slicing, which has no impact on the way the loops are constructed), and will be interpreted as a method for estimating transverse spin correlations in Sec.~\ref{OnotinH}. In Fig.~\ref{fig:Clusterupdate} we similarly illustrate a cluster update for the TFIM, where the spin-flipping operators in the Hamiltonian, along with single-site diagonal operators with the same values of the matrix elements added to $H$, provide the boundaries of clusters that can be flipped independently of each other. These loop and cluster updates greatly reduce autocorrelation times relative to local moves. They are essential for ergodicity in the case of the XXZ model; local Metropolis updates of off-diagonal operators are unable to generate transitions between winding number sectors.

It is well known that the directed-loop update provides a natural framework for measuring off-diagonal correlation functions in XXZ models \cite{Dorneich_PRE_2001} (similar to the worm algorithms formulated in continuous imaginary time \cite{Prokofev_JETPL_1996}). As we will show in Secs.~\ref{DCorr}-\ref{OnotinH}, time-sliced SSE allows better computational efficiency than the original SSE formulation. In the case of the TFIM, the cluster algorithm \cite{Sandvik_PRE_2003} in itself is not directly related to the measurement of off-diagonal single-spin correlation functions, since those operators are present in $H$ regardless of the algorithm used. 

\subsection{Imaginary-time Slicing}

In SSE, imaginary time is implicitly represented by the locations of the ordered operators in the expansion; the index $p \in \{1,\ldots,M\}$ in Eq.~(\ref{Eq:SSEZ}) or, in practice, at the stage of measuring observables, the index $\in \{1,\ldots,n\}$ of the reduced sequence with the unit operators removed.
In conventional SSE formulations without explicit or well-defined imaginary-time points, the evaluation of imaginary-time correlation functions requires a computationally expensive binomial expansion over the operator string~\cite{Sandvik_JPA_1992,Dorneich_PRE_2001}. Though only a relatively narrow range of distances in the string contribute to a given time separation $\tau$, the summations still require considerable effort. A convenient way to avoid this difficulty is to sample the distribution \cite{Pippan_PRA_2009}. Here we exploit another compelling approach by introducing an explicit discretization of the operator string into $m$ ``slices'', where the exponential operator is written as $(e^{-\Delta_\tau H})^m$ and $\Delta_\tau = \beta/m$~\cite{Sandvik2019SSE,Shao_PRB_2015}. The operator string then consists of $m$ segments that are individually expanded to all contributing orders, each with a cutoff $M$ (so that the total operator string length is now $mM$, which is of the same order as $M$ in the method without slicing). The boundaries between these segments correspond to well-defined imaginary-time values for which the correlation functions can be computed rapidly.

The sliced partition function is
\begin{equation}
    Z = \text{Tr}\left[\exp{\left(-\beta H\right)}\right] = \text{Tr}\left[\exp{\left(-\Delta_\tau H\right)^m}\right].
\end{equation}
Instead of performing a single series expansion in $\beta$ as before, we now expand the exponential in each time slice $l$ independently
in the slice width $\Delta_\tau$. Let $n_l$ be the number of non-identity local operators $H_b$ in slice $l$. Then for each slice,
\begin{equation}
    e^{-\Delta_\tau H} = \sum_{n_{l}=0}^{\infty} \frac{(-\Delta_\tau)^{n_{l}}}{n_{l}!}H^{n_{l}}
\end{equation}
so that we can write the partition function as
\begin{equation}
    Z = \sum_{\mathcal{C}}\prod_{l=0}^{m-1}\left[\frac{(\Delta_\tau)^{n_l}}{n_l!}\right]\expec{\alpha}{\prod_{l}\mathcal{S}_{l}}{\alpha},
\end{equation}
where $\mathcal{S}_{l}$ denotes the operator string in slice $l$ and we have defined the sum over all SSE configurations as 
\begin{equation}
    \sum_{\mathcal{C}} \equiv \sum_{\{\alpha\}}\sum_{\{n_l\}}\sum_{\{\mathcal{S}_{n_l}\}}.
\end{equation}

This construction is equivalent to the conventional non-sliced form, but with the added benefit that we now have definite imaginary-time coordinates $\tau_k=k\Delta_\tau$ for $k \in \{0,\dots,m-1\}$ defined on the boundaries between each time slice. In practice, a cutoff is introduced as in Eq.~\eqref{Eq:SSEZ} so that each string segment is of the length $M$. Estimators of observables can still be derived using a formulation with no cutoff, as we will do below for simplicity. This simply corresponds to neglecting the unit operators in the string.

The sliced formulation does not significantly change the SSE algorithm. We can still perform sampling consisting of diagonal and directed-loop or cluster updates. The only significant (but very simple) difference is in the diagonal update, where the insertion and removal probabilities depend on the time-slice index $k$ through the number of non-unit operators $n_l$ within the slice. The probabilities to add or remove a local operator $H_b$ in place of an identity in time slice $k$ are
\begin{subequations}\label{slicedp}
\begin{align} 
    P(\mathbb{I}\rightarrow H_b) &= \frac{N_b\Delta_\tau\left[H_b\right]}{M-n_k}, \\
    P(H_b\rightarrow \mathbb{I}) &= \frac{M-n_k+1}{N_b\Delta_\tau\left[H_b\right]}.
\end{align}
\end{subequations}

Compared with Eqs.~\eqref{eq:probs}, we have replaced $\beta$ with $\Delta_\tau$, and $n$, the number of bond operators in the entire string, with $n_k$, the number in only time slice $k$. 

The directed loops for the XXZ model and the cluster update for the TFIM do not change at all, nor do any measurements of the standard observables. The slicing does not bias loop or cluster updates since loop construction is local to vertices and independent of slice boundaries.  With the sliced representation, we can measure correlation functions on the set of well-defined $\tau_k$ points. There are, however, some subtleties that require close attention in the case of correlations of off-diagonal operators and working out the correct expressions and the corresponding unbiased computational procedures is the main aim of the work presented here.

One thing to note here is that a small time slice width $\Delta_\tau$ leads to large variance of the numbers $n_l$ of Hamiltonian operators in the slices, thus necessitating a large cut-off $M$, in extreme cases $M \gg \langle n_l\rangle$. The total string length $mM$ can then be significantly larger (though still scaling in the same way) as just $M$ for the case of a single slice ($\Delta_\tau=\beta$). This minor drawback diminishes with increasing system size $N$, because the number of operators in the slices also scales with $N$. In practice, the somewhat larger string length with slicing is therefore not a serious issue, significantly affecting only simulations with small $N$ and/or very small $\Delta_\tau$.

\subsection{Imaginary-time correlation functions}

We can write a general imaginary-time correlation function as
\begin{equation}
    G_{\mathcal{O}}(r,\tau_k) = \bracket{\mathcal{O}_1(r_0+r, \tau_{k_0}+\tau_k)\mathcal{O}_2(r_0, \tau_{k_0})},
\end{equation}
at separation $(r,\tau_k)$, with $\tau_k \equiv k\Delta_\tau$, $k \in \{0,1,\ldots,m-1\}$ being one of the accessible time separations given the slice width $\Delta_\tau$. Note that we label the time points and separations starting from zero index, so the boundary $\tau_k$ corresponds to the state generated from the reference state $\ket{\alpha}$ by the operators in the first $k$ slices. In simulations, the reference points $r_0,\tau_{k_0}$ will typically be averaged over, taking into account also the periodic imaginary-time and (normally) space boundary conditions. Without loss of generality, in the discussion below we will take the initial coordinates $(r_0, \tau_{k_0}) = (0,0)$ for simplicity in our expressions.

We now write the operators in the Heisenberg picture,
\begin{equation}
    \begin{aligned}
        G_{\mathcal{O}}(r, \tau_k)=\frac{1}{Z}\text{Tr}\left[e^{-(\beta-\tau_k)H}\mathcal{O}_1(r)e^{-\tau_k H}\mathcal{O}_2(0)\right]
\end{aligned}
\end{equation}
and adapt the previous derivation of the SSE partition function. With $\beta = m\Delta_\tau$, we can expand each time-slice as its own operator string,
\begin{equation}\label{eq:gencorrelator}
\begin{aligned}
    G_{\mathcal{O}}&(r, \tau_k)= \frac{1}{Z}\sum_\mathcal{C}\left[\prod_l\frac{(\Delta_\tau)^{n_l}}{n_l!}\right]\\&\times\expec{\alpha}{\left(\prod_{l\geq k}\mathcal{S}_l\right)\mathcal{O}_1(r)\left(\prod_{l<k}\mathcal{S}_l\right)\mathcal{O}_2(0)}{\alpha}
\end{aligned}
\end{equation}

The operators $\mathcal{O}_1$ and $\mathcal{O}_2$ are now separated by $k$ time-sliced operator strings; thus, they have well defined imaginary times, in contrast to operators placed within time slices (as in the original SSE with a single slice $\Delta_\tau=\beta$). By cyclicity of the trace, we may permute any number of operator strings $\mathcal{S}_l$ such that we can place $\tau_{k_0}$ at any desired imaginary-time location and the choice of $\tau_{k_0} = 0$ was justified. 

\section{Diagonal Correlation Functions}\label{DCorr}

We can directly access the full time-propagated state $\ket{\alpha(\tau_k)}$ at each time-slice boundary, and so correlators of diagonal operators at well defined imaginary-time points can be obtained straightforwardly by evaluating operator eigenvalues at these boundaries, as discussed previously in Ref.~\cite{Sandvik2019SSE} and already applied at large scale in several works, e.g., \cite{Shao_PRX_2017,Qin_PRB_2017,Ma_PRB_2018,Yang_PRB_2025}. We here briefly summarize this standard construction to fix notation and to contrast with the off-diagonal case discussed in Secs.~\ref{OinH} and \ref{OnotinH}.

Continuing from Eq.~(\ref{eq:gencorrelator}), the operators $\mathcal{O}_1$ and $\mathcal{O}_2$ may be replaced by their eigenvalues at the corresponding points in the operator string. If $\ket{\alpha(\tau_k)}$ denotes the propagated spin state at time-slice boundary $\tau=k\Delta_\tau$, $k\in\{0,\ldots,m-1\}$, then
\begin{equation}
    \left[\mathcal{O}(r,\tau_k)\right] = \expec{\alpha(\tau_k)}{\mathcal{O}(r)}{\alpha(\tau_k)}
\end{equation}
is simply the eigenvalue of the diagonal operator acting on that state. We can pull these values out of the operator string to get
\begin{equation}
\begin{aligned}
    G_{\mathcal{O}}&(r, \tau_k)= \frac{1}{Z}\sum_\mathcal{C}\left[\prod_l\frac{(\Delta_\tau)^{n_l}}{n_l!}\right]\times\\&\expec{\alpha}{\left(\prod_{l}\mathcal{S}_l\right)}{\alpha}\left[\mathcal{O}_1(r,\tau_k)\right]\left[\mathcal{O}_2(0,0)\right].
    \end{aligned}
\end{equation}
This expression has the standard form of a Monte Carlo estimator in the SSE framework, consisting of a sum over configurations weighted by their statistical weights and an associated estimator function. The correlator thus reduces to an average of products of eigenvalues over Monte Carlo configurations. Explicitly, we can write
\begin{equation}
    G_{\mathcal{O}}(r, \tau_k) = \frac{\sum_\mathcal{C}W(\mathcal{C})f(r,\tau_{k};\mathcal{C})}{\sum_\mathcal{C}W(\mathcal{C})},
    \label{eq:estimator}
\end{equation}
where
\begin{equation}
    W(\mathcal{C}) = \left[\prod_l\frac{(\Delta_\tau)^{n_l}}{n_l!}\right]\expec{\alpha}{\left(\prod_{l}\mathcal{S}_l\right)}{\alpha}
\end{equation}
is the sampling weight of the configuration $\mathcal{C}$, with the fill-in unit operators used in the acceptance probabilities Eqs.~(\ref{slicedp}) now disregarded, such that $Z = \sum_\mathcal{C}W(\mathcal{C})$ and 
\begin{equation}
    f(r,\tau_{k};\mathcal{C}) = \left[\mathcal{O}_1(r,\tau_k)\right]\left[\mathcal{O}_2(0,0)\right]
\end{equation}
is the estimator that we accumulate. Thus, we evaluate
\begin{equation}
    \bracket{G_{\mathcal{O}}(r, \tau_k)} = \bracket{f(r,\tau_{k};\mathcal{C})}_W,
\end{equation}
where the $W$ subscript denotes that the average is taken over configurations sampled with weights $W(\mathcal{C})$. 

In practice, one records the values of $\mathcal{O}(r,\tau_k)$ in the computational basis at all time-slice boundaries as the operator string is traversed. This yields access to the observable on an $m\times L^d$ grid for a system in $d$ spatial dimensions. A naive evaluation of the full space-time averaged correlator $G_\mathcal{O}(r,\tau_k)$ from these data would then scale as $\mathcal{O}(m^2L^{2d})$. Often the correlation functions are only needed along some specific lines, however, and the computational effort is then greatly reduced. 

When the full space-time correlator $G_\mathcal{O}( r,\tau_k)$ is required (rather than a small subset of all spatial separations), it is more efficient to use a fast-Fourier-Transform (FFT) approach. Let us define
\begin{equation}
    \mathcal{O}(r,\tau_k) = \frac{1}{mN}\sum_{q,\omega_n}\tilde{\mathcal{O}}(q,\omega_n)e^{i(\omega_n\tau_k + qr)},
\end{equation}
where the momentum $q = 2\pi n/N$ (in one dimension, with simple generalization to any dimensionality), $n=0,\ldots,N-1$ and $\omega_n$ are the Matsubara frequencies; $\omega_n = 2n\pi T$, $n=0,\ldots,m-1$ for bosonic operators (this set also being finite because of the time discretization). This allows us to rewrite the estimator as
\begin{equation}
    f({r,\tau_k}) = \frac{1}{mN}\sum_{q,\omega_n}|\tilde{\mathcal{O}_1}(q,\omega_n)\mathcal|^2e^{i(\omega_n\tau_k + qr)}
\end{equation}
where computationally the FFT and its inverse transform IFFT are used (note we've set $\mathcal{O}_1=\mathcal{O}_2$ for simplicity);
\begin{equation}
    \bracket{f({r,\tau_k})}_W = \bracket{\text{IFFT}\left[\abs{\text{FFT}\left[\mathcal{O}(r,\tau_k)\right]}^2\right]}_W.
\end{equation}
This procedure scales as $\mathcal{O}(mL^d\log{(mL^d)})$; a significant speedup. For the often needed momentum resolved correlator $G_\mathcal{O}(q,\tau_k)$ (in analytic continuation based on the $\tau$ axis instead of the $\omega_n$ space \cite{ShaoSAC}), the IFFT is taken only in the Matsubara dimension.  

In many cases, loop and cluster algorithms also allow for improved estimators, which are obtained by summing analytically over all cluster or loop ``orientations''---for the $S = 1/2$ model considered here the two Ising-like states of each cluster or loop. In the case of time sliced SSE, cluster or loop labels are then stored for each time slice at the stage of decomposing the space-time lattice into loops/clusters. Improved estimators can often dramatically increase the statistical precision of correlation functions, but unfortunately it is then not possible to efficiently implement the FFT trick. Whichever approach is more efficient in practice will depend on the model and the quantities of interest. 

\section{Off-diagonal correlation functions with $\mathcal{O} \in H$}\label{OinH}

We now turn to the measurement of imaginary-time correlation functions of off-diagonal operators. These fall into two cases that must be treated separately: operators contained in the Hamiltonian, and those that are not. We treat the first case in this section and the second in Sec.~\ref{OnotinH}. A simple example of the first case is the correlator $\bracket{\sigma^x(r,\tau)\sigma^x(r',\tau')}$ in the TFIM (Sec.~\ref{sec:models}), where $\sigma^x$ appears explicitly in the Hamiltonian. 

Before we discuss how to measure such correlation functions, we briefly review the more basic example of measuring the expectation value of an operator $\bracket{H_k}$ with $H_k\in H$. Following the notation of Sec.~3 of Ref.~\cite{Sandvik_JPA_1992},
\begin{equation}
     \bracket{H_k} = \frac{1}{Z}\sum_{\alpha}\sum_{n=0}^\infty\sum_{S_n}\frac{(-\beta)^n}{n!}
       \expec{\alpha}{H_k\prod_{i=1}^nH_{l_i}}{\alpha}.
\end{equation}
Absorbing the additional $H_k$ into the operator string and reindexing $n \to n+1$, we can interpret this as a longer operator string that contains $H_k$ as its final element
\begin{equation}\label{Eq:Reindex1}
\begin{aligned}
     \bracket{H_k} = \frac{1}{Z}\sum_{\alpha}\sum_{n=1}^\infty\sum_{S_{n}}\frac{(-\beta)^{n-1}}{(n-1)!}
       \expec{\alpha}{\prod_{i=1}^{n}H_{l_i}}{\alpha}
       \end{aligned}
\end{equation}
where $H_{l_{n}} = H_k$. Reworking the weight by explicitly factoring out $-n/\beta$ and extending the sum down to $n=0$, we can write
\begin{equation}\label{Eq:Reindex2}
     \bracket{H_k} = \frac{1}{Z}\sum_{\alpha}\sum_{n=0}^\infty\sum_{S_{n}}\left(-\frac{n}{\beta}\right)
       \frac{(-\beta)^{n}}{n!}\expec{\alpha}{\prod_{i=1}^{n}H_{l_i}}{\alpha}.
\end{equation}
Recognizing the weights $W(\mathcal{C})$ as in Eq.~(\ref{eq:estimator}), we can define the MC estimator
\begin{equation}
    f(\mathcal{C}) =
\begin{cases}
  -n/\beta, & \text{final operator in the string is $H_k$}, \\
  0, & \text{otherwise}.
\end{cases}
\end{equation}
In practice, we average this over the entire operator string and obtain the familiar result
\begin{equation}\label{eq:expecHk}
    \bracket{H_k} = -\frac{1}{\beta}\bracket{N_k}, 
\end{equation}
where $N_k$ is simply the count of operators $H_k$ appearing in the string. The key observation is that for operators $\mathcal{O}\in H$, expectation values (and also correlation functions) can be obtained by counting instances of operators (or pairs of operators) in the string. This again comes from interpreting the observable as a contribution to the partition function with an additional weight factor. 

One of the simplest applications of this estimator is the measurement of the internal energy $E = \bracket{H}$. This corresponds to counting all $H_k \in H$, and yields the estimator
\begin{equation}
    E = \bracket{H} = -\frac{1}{\beta}\bracket{n},
\end{equation}
where $n$ is just the total number of non-identity operators in the operator string. 

Now we will compute the correlation function of two operators $\mathcal{O}_1, \mathcal{O}_2\in H$ using Eq.~(\ref{eq:gencorrelator}). For simplicity, let us write the time-slice series expansion factors as
\begin{equation}
    w(\mathcal{C}) = \prod_l\frac{(\Delta_\tau)^{n_l}}{n_l!}
\end{equation}
and then expand out the expression 
\begin{widetext}
    \begin{equation}
        \begin{aligned}\label{Eq:ODinHLong}\bracket{\mathcal{O}_1(r,\tau_k)\mathcal{O}_2(r',\tau_{k'})}=\frac{1}{Z}\sum_\mathcal{C}w(\mathcal{C})\bigg\langle\alpha\bigg|\left(\prod_{l=k+1}^{m-1}\mathcal{S}_l\right)\left(H_{k,n_k}\cdots H_{k,1}\right)\mathcal{O}_1(r)\left(H_{k-1,n_{k-1}}\cdots H_{k-1,1}\right)\times\\\left(\prod_{l=k'+1}^{k-2}\mathcal{S}_l\right)\left(H_{k',n_{k'}}\cdots H_{k',1}\right)\mathcal{O}_2(r')\left(H_{k'-1,n_{k'-1}}\cdots H_{k'-1,1}\right)\left(\prod_{l=0}^{k'-2}\mathcal{S}_l\right)\bigg|\alpha\bigg\rangle.
    \end{aligned}
    \end{equation}
\end{widetext}

Here, we have explicitly expanded the operator strings to the left and right of the two $\mathcal{O}$ insertions. The indices of each $H_{k,i}$ denotes the local Hamiltonian operator at position $i$ in time slice $k$. Note that $\mathcal{O}_2$ has been inserted at time slice $k'$ not necessarily equal to zero. This is purely for notation, and by cyclicity of the trace we can move $\tau_{k'}$ anywhere.

\begin{figure}[t] 
  \centering
  \includegraphics[width=\columnwidth]{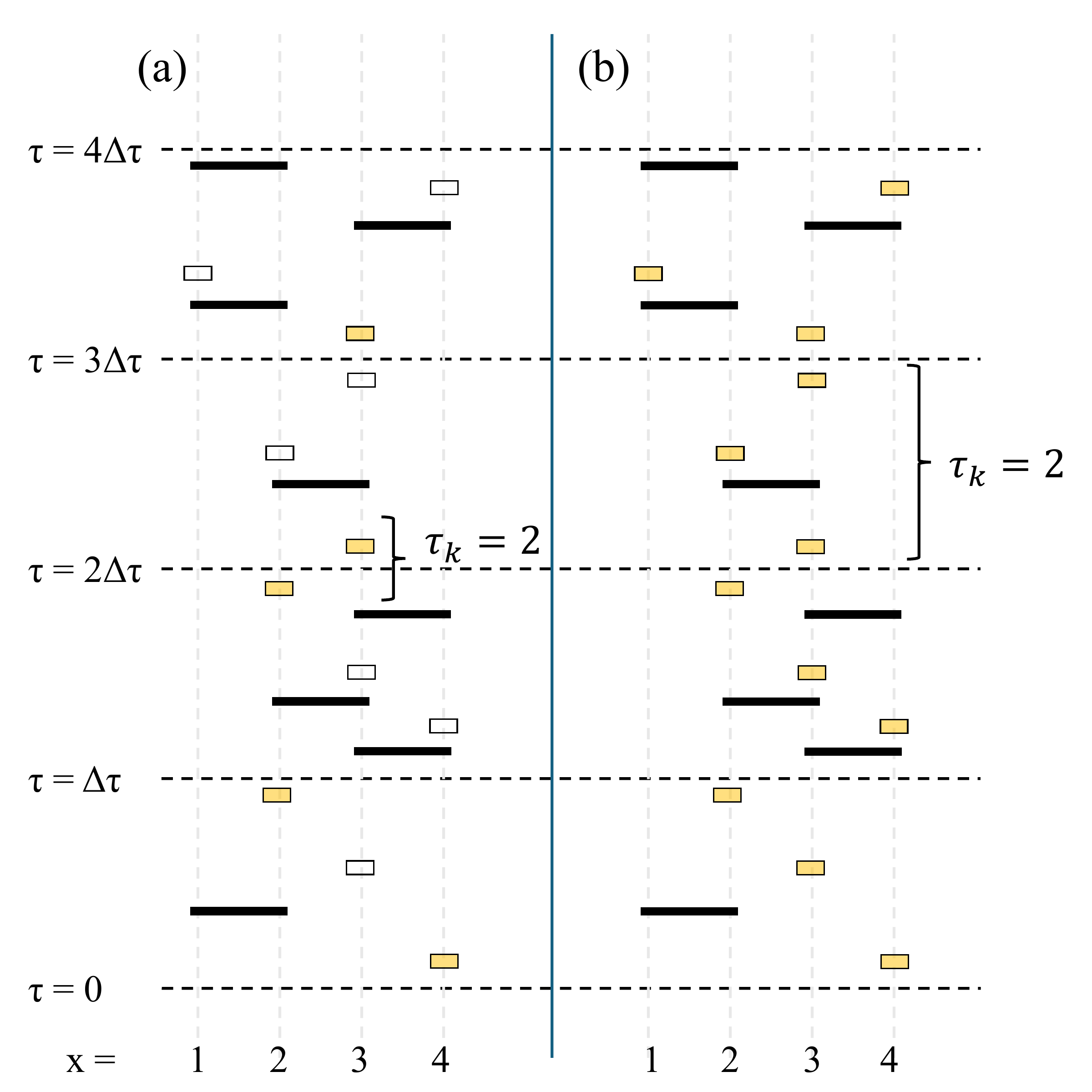}
  \caption{Illustration of the off-diagonal estimator for the correlation function $\bracket{\sigma^x(r,\tau)\sigma^x(0,0)}$ in the TFIM, using the (a) boundary-valued and the (b) time-slice averaged formulations. Two-site Ising $J$ operators are denoted by black rectangles, and do not factor into the measurement procedures. Diagonal single-site operators $h\mathbb{I}$ are not shown, and also do not factor into the procedure. Spin-flip operators are denoted by single-site rectangles, where yellow denotes that they contribute to the estimator. For the (a) boundary-valued estimator, only spin-flip operators directly adjacent to the time-slice boundary contribute. Only a fraction of total operators are used. This method is exact, but leads to higher variance than methods counting all operators. For the (b) time-slice averaged estimator, all spin-flip operators contribute. However, there is a bias of $\mathcal{O}(\Delta_\tau^2)$.}
  \label{fig:OinH}
\end{figure}

We now apply the same logic as we did in the simpler $\bracket{H_k}$ case above. The correlation function can be viewed as a configuration in which each $\mathcal{O}_i$ is absorbed into an adjacent time-slice string: as the first operator in the string to the left, or equivalently as the last operator in the string to the right (reading right to left). Insertion into slice $l$ will introduce a factor of $n_l/\Delta_\tau$ analogous to the factor of $-n/\beta$ in the single-operator case. This follows from the same reindexing argument as in Eqs.~(\ref{Eq:Reindex1})-(\ref{Eq:Reindex2}), now applied within a single time slice. 

Therefore, for each operator $\mathcal{O}_i(\tau_k)$, we can imagine inserting it into slice $k-1$ (to the right) or slice $k$ (to the left) and producing a factor of $n_{k-1}/\Delta_\tau$ or $n_{k}/\Delta_\tau$ respectively. The estimator is then the product of the two factors for $\mathcal{O}_1(\tau_k)$ and $\mathcal{O}_2(\tau_{k'})$. Explicitly, for $\Delta\tau_k\equiv\tau_k-\tau_{k'}\neq 0$ (the case $\Delta\tau_k=0$ is special and will be considered later):
\begin{equation}\label{eq:f-dtne0}
f_{r,\Delta\tau_k\neq 0}(\mathcal{C}) =
\begin{cases}
\dfrac{n_l n_{l'}}{\Delta_\tau^2}, & \parbox[t]{0.5\linewidth}{if $l$ and $l'$ correspond to the slices adjacent to $\mathcal{O}_1$ and $\mathcal{O}_2$,} \\[15pt]
0, & \text{otherwise.}
\end{cases}
\end{equation}

Then $l$ and $l'$ are defined as follows, referring to the notation in Eq.~(\ref{Eq:ODinHLong}): if $\mathcal{O}_1 = H_{k,1}$,
then $l = k$, while if $\mathcal{O}_1 = H_{k-1,n_{k-1}}$ then $l = k-1$.
Similarly, if $\mathcal{O}_2 = H_{k',1}$ then $l' = k'$, and if
$\mathcal{O}_2 = H_{k'-1,n_{k'-1}}$ then $l' = k'-1$.

In the $\bracket{H_k}$ calculation, we invoked cyclicity of the trace to average over all possible insertion slots, dividing out the $n$ factor. Here, $\mathcal{O}(r,\tau_k)$ has a definite imaginary-time coordinate and we now use the integer index $k$ as a more convenient way to address a particular slice. While all time-slice boundaries are equivalent by cyclicity, they are distinct from the slice bulks in which $\tau$ is not well defined, and we cannot simply average over all possible positions as before (without incurring some error in $\Delta_\tau$, which we do consider below). The estimator therefore depends on the configuration-dependent $n_l$, the number of non-identity bond operators in slice $l$. This also means the estimator will have higher variance, since we cannot average over the entire operator string, and more measurements are necessary to achieve error bars comparable to the diagonal case. 

In practice, the algorithm for computing this estimator can be written as follows. Let $N$ be the number of lattice sites, $m$ the number of the time-slice boundaries, and $r$ denote the site index.

\par\bigskip
\noindent\hrule height 0.8pt
\vspace{3pt}
\begin{algorithmic}[]
    \State Define $N_\mathcal{O} \gets \text{zero array of size } (N, m)$
    \State \% Loop through the operator string
    \For{each boundary $k$}
    \If{last operator in slice $k-1$ is $\mathcal{O}(r)$}
        \State $N_\mathcal{O}[r,k] = n_{k-1}$
    \ElsIf{first operator in slice $k$ is $\mathcal{O}(r)$}
        \State $N_\mathcal{O}[r,k] = n_k$
    \Else
        \State $N_\mathcal{O}[r,k] = 0$
    \EndIf
    \EndFor
\end{algorithmic}
\vspace{3pt}
\noindent\hrule height 0.8pt
\bigskip

Here the variable $N_\mathcal{O}[r,k]$ holds the information necessary to compute the correlator. Note that we avoid double counting operators at any given time-slice boundary by only counting $\mathcal{O}(r)$ in slice $k-1$ or $k$, but not both. This counting procedure occurs during the normal measurement step in SSE, where we already step through the operator string, and therefore introduces no additional computational scaling, with memory overhead limited to the $N_\mathcal{O}[r,k]$ array. An illustration of this algorithm is also shown in Fig.~\ref{fig:OinH}(a). Note that at the boundary $\tau = 3\Delta\tau$, only one $\sigma^x(r=3)$ operator contributes to the estimator to avoid double counting. One could also count both for slightly better statistics, but would need to be careful about factors of $2$.

We can now compute $G_\mathcal{O}(r,\tau_k)$ by summing all pairwise contributions from $N_\mathcal{O}[r,k]$. For full space-time resolution, this would scale as $\mathcal{O}(m^2L^{2d})$ as in the diagonal case. We can again apply the FFT for a more efficient computation. The final result will be correct up to the multiplicative factors of $\Delta_\tau^2$ and normalization, which can be reintroduced trivially.

The above algorithm correctly computes the full correlation function for all $\Delta\tau_k > 0$, but for the equal-time correlator we must slightly alter the procedure. For $\Delta\tau_k = 0$, the two operators must occupy adjacent positions on opposite sides of the boundary $\tau_k$: $\mathcal{O}_1$ as the first operator of slice $k$ and $\mathcal{O}_2$ as the last operator of slice $k-1$. The estimator is then
\begin{equation}
f_{r, \Delta\tau_k=0}(\mathcal{C}) =
   \begin{cases}
   \dfrac{n_k n_{k-1}}{\Delta_\tau^2}, & \parbox[t]{0.3\linewidth}{if $\mathcal{O}_1 = H_{k,1}$ and $\mathcal{O}_2 = H_{k-1,n_{k-1}}$,} \\[15pt]
   0, & \text{otherwise.}
   \end{cases}
\end{equation}

This calculation can be performed at the same time as the general $\Delta\tau_k \neq 0$ case above, and the two combined to give the full correlation function. 

We often need the \emph{connected} correlator $G_{\mathcal{O}}^{\text{conn}}(r,\tau) = \bracket{\mathcal{O}_1(r,\tau)\mathcal{O}_2(0,0)} - \bracket{\mathcal{O}_1}\bracket{\mathcal{O}_2}$. To do this, simply accumulate $\bracket{\mathcal{O}_i}$ using Eq.~(\ref{eq:expecHk}), which amounts to counting all instances of the operator in the string, and subtract from the full correlator. 

Note that for the TFIM that $H_k = h\sigma^x$, with the coupling $h$ included in the operator. Therefore, when we specifically want to compute the correlation function $G^{xx}(r,\tau) = \langle \sigma^x(r,\tau)\sigma^x(0,0)\rangle$, we need to divide the estimator by an additional factor of $h^2$. This holds in general for any correlator built from counting operators, where if the operator is $g\mathcal{O}$ one must divide out $g^2$ to obtain the correct correlations. The statistical errors will therefore be large for small $g$, where the sampling itself also is slow. In practice, the most interesting cases do not involve very small $g$, however, e.g., in studies of quantum phase transitions.

\subsection{Improved Estimator and Bias-Variance Tradeoff}

As mentioned above, the constraint that operators can only be counted adjacent to the boundaries (and not averaged over the entire string) leads to a higher variance estimator than in the diagonal correlation functions. This constraint is necessary to evaluate the off-diagonal correlations with no error. However, in some cases it may be worthwhile to instead compute a time-slice averaged estimator, which introduces an error of $\mathcal{O}(\Delta_\tau^2)$. An illustration of this algorithm is shown in Fig.~\ref{fig:OinH}(b). In contrast to the boundary-valued estimator, all spin-flip operators in the string contribute, and thus there can be a significant reduction in the variance. 

Instead of counting operators $\mathcal{O}_i$ only on the boundaries $\tau_k$, let us instead introduce the time-slice averaged correlation function,
\begin{equation}
\begin{aligned}
    \bar{G}_\mathcal{O}&(r,\tau_k) = \frac{1}{Z}\sum_{\mathcal{C}}w(\mathcal{C})\frac{1}{n_k+1}\frac{1}{n_0+1}\\
    &\times\sum_{p=0}^{n_k}\sum_{q=0}^{n_0}\expec{\alpha}{\left(\prod_{l>k}\mathcal{S}_l\right)\mathcal{S}_k^p\left(\prod_{l<k}\mathcal{S}_l\right)\mathcal{S}_0^q}{\alpha}
\end{aligned}
\end{equation}
where we now average over all insertion locations for $\mathcal{O}_1$ and $\mathcal{O}_2$ in the time-slices $\tau_k$ and $\tau_0$ respectively. $\mathcal{S}_k^p$ denotes the operator string for time slice $k$ with $\mathcal{O}_i$ inserted at position $p$. $\mathcal{O}_1$ is inserted in slice $\mathcal{S}_k$, while $\mathcal{O}_2$ is  in $\mathcal{S}_0$. In the above expression we then apply the same operator-absorption argument, re-indexing and factoring out the weights, the estimator (for non-zero time-slice separations $\tau_k\neq 0$) is
\begin{equation}\label{Eq:IEdiffT}
    \bar{G}_\mathcal{O}(r,\tau_k\neq 0) =\frac{1}{\Delta_\tau^2} \bracket{N_{\mathcal{O}_1}(r,\tau_k)N_{\mathcal{O}_2}(0,0)},
\end{equation}
with $N_\mathcal{O_i}(r,\tau_k)$ the number of operators $\mathcal{O}_i$ appearing at site $r$ in time slice $\tau_k$.

For same-time correlations $\tau_k=0$, the estimator is again slightly different. We need to account for the ordering within the same slice, so we only accumulate the upper triangular contributions
\begin{equation}\label{Eq:IEsameT}
    \bar{G}_\mathcal{O}(r,0) = \frac{1}{2\Delta_\tau^2}\bracket{N_\mathcal{O}(r,0)\left(N_\mathcal{O}(0,0)-\delta_{r,0}\right)}.
\end{equation}
These two estimators can be combined easily using the FFT trick. The FFT computes, for each separation $(r,\tau_k)$, the sum of pairwise products $N_\mathcal{O}(r'+r,\tau_k'+\tau_k)N_\mathcal{O}(r',\tau_k')$ over all sites and time slices. Let
\begin{equation}
    C(r,\tau_k) = \frac{1}{\Delta_\tau^2}\text{IFFT}\left[\abs{\text{FFT}\left[N_\mathcal{O}(r,\tau_k)\right]}^2\right]
\end{equation}
For non-zero separations, $C(r,\tau_k)$ is exactly the correct estimator in Eq.~\eqref{Eq:IEdiffT}. However, at $(r,\tau_k) = (0,0)$, the FFT pairs each operator count with itself, yielding $\sum_{r,\tau_k}N_\mathcal{O}(r,\tau_k)^2/\Delta_\tau^2$. The correct equal-time estimator requires only distinct pairs within the same site and time slice, proportional to $N_\mathcal{O}(N_\mathcal{O}-1)$. The difference between these two is exactly $\sum_{r,\tau_k}N_\mathcal{O}(r,\tau_k)$, so we correct the same-site equal-time entry by subtracting this self-correlation:
\begin{equation}
    \tilde{C}(0,0) = C(0,0)-\frac{1}{\Delta_\tau^2}\sum_{r,\tau_k}N_\mathcal{O}(r,\tau_k),
\end{equation}
and average to obtain the Green's function $\bar{G}_\mathcal{O}(r,\tau_k) = \langle\tilde{C}(r,\tau_k)\rangle$. 

There is an error associated with averaging insertion over the entire time slice that does not appear in the boundary estimator. We are considering insertions at arbitrary (not discrete) imaginary times such that $\tau = k\Delta_\tau + \Delta_s$ for $\Delta_s\in(0,\Delta_\tau)$ uniformly distributed within the time slice. The average insertion time is therefore
\begin{equation}
    \bracket{\tau} = k\Delta_\tau + \Delta_\tau/2.
\end{equation}
However, since we are only interested in separations, the average imaginary-time separation is
\begin{equation}
    \bracket{\delta\tau} = (k-k')\Delta_\tau + \bracket{\Delta_{s_1}} - \bracket{\Delta_{s_2}} = (k-k')\Delta_\tau.
\end{equation}
In the average, the uniformly distributed $\Delta_{s_i}$ cancel and we get the correct separation. Now, the variance of the uniform distribution is $\bracket{\Delta_s^2} = \Delta_\tau^2/12$. Therefore the variance in the imaginary-time separation (which will be the source of the error in the correlation function) is
\begin{equation}
    \bracket{\delta\tau^2} = 2\bracket{\Delta_s^2} = \Delta_\tau^2/6.
\end{equation}
We can then derive the expected error in the computed correlation function to lowest order in $\Delta_\tau$
\begin{equation}
\begin{aligned}
    \bar{G}(\tau_k) &= \bracket{G(k\Delta_\tau+(\Delta_{s_1}-\Delta_{s_2}))}_{\Delta_s}\\
    &\approx \bracket{G(\tau_k)} + \frac{\Delta_\tau^2}{12}\bracket{G''(\tau_k)} + \ldots.
\end{aligned}
\end{equation}

The error in the estimator scales as $\mathcal{O}(\Delta_\tau^2)$, although the dependence on the second derivative $G''$ is also important. At $\tau_k=0$, the correlation function has a cusp due to the imaginary-time periodicity and the error instead scales with $\mathcal{O}(\Delta_\tau)$. 

If we are only interested in long-time correlations, $\tau \gg \Delta_\tau$, which is often the case as these control the low-frequency excitations, this larger error in early times is not an important factor. Additionally, since we now average over all operators in the string (instead of only at the boundaries), the variance of the estimator can be significantly reduced. This leads to a bias-variance tradeoff. For larger $\Delta_\tau$, there are more operators within each slice, and the variance reduction is more significant. However, this is also the regime in which the error is largest. For very small $\Delta_\tau$, the average number of operators within each slice approaches $\lesssim 1$, and the time-slice averaged estimator reduces exactly to the boundary estimator. In practice then, the ``better'' method will depend on both the model and the chosen time-slice resolution $\Delta_\tau$. For the purpose of just studying the decay of correlations in imaginary time, the slice-density approach will be precise enough. In fact, as we will demonstrate below, the errors are surprisingly small even with rather large $\Delta_\tau$. Still, if the results will be used in analytic continuation some care should be taken to test the dependence of final results on $\Delta_\tau$.

\section{Off-Diagonal Correlation Functions with $\mathcal{O} \notin H$}\label{OnotinH}

We finally consider the case of correlation functions of operators that are off-diagonal in the computational basis and do not appear explicitly in the Hamiltonian. A representative example is the correlator $\bracket{S^+(r,\tau) S^-(0,0)}$ in the Heisenberg XXZ model. For concreteness, we work here with spin operators, although the same construction applies directly to bosonic correlators such as $\bracket{ c^\dagger(r,\tau)c(0,0)}$.

It is well established that such off-diagonal correlation functions can be accessed in QMC simulations by introducing defects in the world-line configuration, as in worm-type or directed-loop algorithms~\cite{Prokofev_JETPL_1996,Dorneich_PRE_2001,Syljuasen_PRE_2002}. In these approaches, the insertion and propagation of a pair of defects naturally samples off-diagonal operator correlations. However, in conventional SSE formulations these estimators are not naturally formulated at fixed imaginary-time separations, making the extraction of $\tau$-resolved correlators computationally costly.

In the following, we adapt these worm and directed-loop ideas to the time-sliced SSE framework. This allows us to construct a direct estimator for off-diagonal correlation functions at well-defined imaginary-time points, combining the advantages of worm-type sampling with the explicit imaginary-time resolution provided by time slicing. As with the $\mathcal{O}\in H$ case above, this will amount to added scaffolding on the loop update, but no additional computational scaling. Specifically, we will compute 
\begin{equation}
    G_\pm(r,\tau_k) \equiv\bracket{S^+(r,\tau_k)S^-(0,0)}.
\end{equation}
For a model with SU(2) symmetry, e.g., the isotropic Heisenberg antiferromagnet, we have
\begin{equation}
\bracket{S^+(r,\tau_k)S^-(0,0)} = 2\bracket{S^z(r,\tau_k)S^z(0,0)}.
\end{equation}
However, as we move away from that isotropic point by tuning either $\Delta$ or $h$, this equality breaks down and $G_\pm(r,\tau_k)$ becomes an independent observable.

We now rewrite Eq.~(\ref{eq:gencorrelator}), explicitly inserting the spin-flip operators
\begin{equation}\label{Eq:ODCorrelator}
\begin{aligned}
G_{\pm}&(r,\tau_k)
    = \frac{1}{Z}
       \sum_\mathcal{C}
       w(\mathcal{C}) \\
    &\quad \times
       \expec{\alpha}{
         \left(\prod_{l \ge k} \mathcal{S}_l\right) S^{+}(r)
         \left(\prod_{l < k} \mathcal{S}_l\right) S^{-}(0)
       }{\alpha} .
\end{aligned}
\end{equation}

In contrast to the diagonal case, the operators $S^\pm$ cannot be replaced by eigenvalues, since their insertion changes the propagated state. Likewise, the counting approach of Sec.~\ref{OinH} does not apply here, as $S^\pm \notin H$ (they only come in pairs $S^+_iS^-_{i+1}$). This is conventionally handled by introducing configurations with open world-line defects, as in worm-type or directed-loop formulations.

\begin{figure}[t] 
  \centering
  \includegraphics[width=0.8\columnwidth]{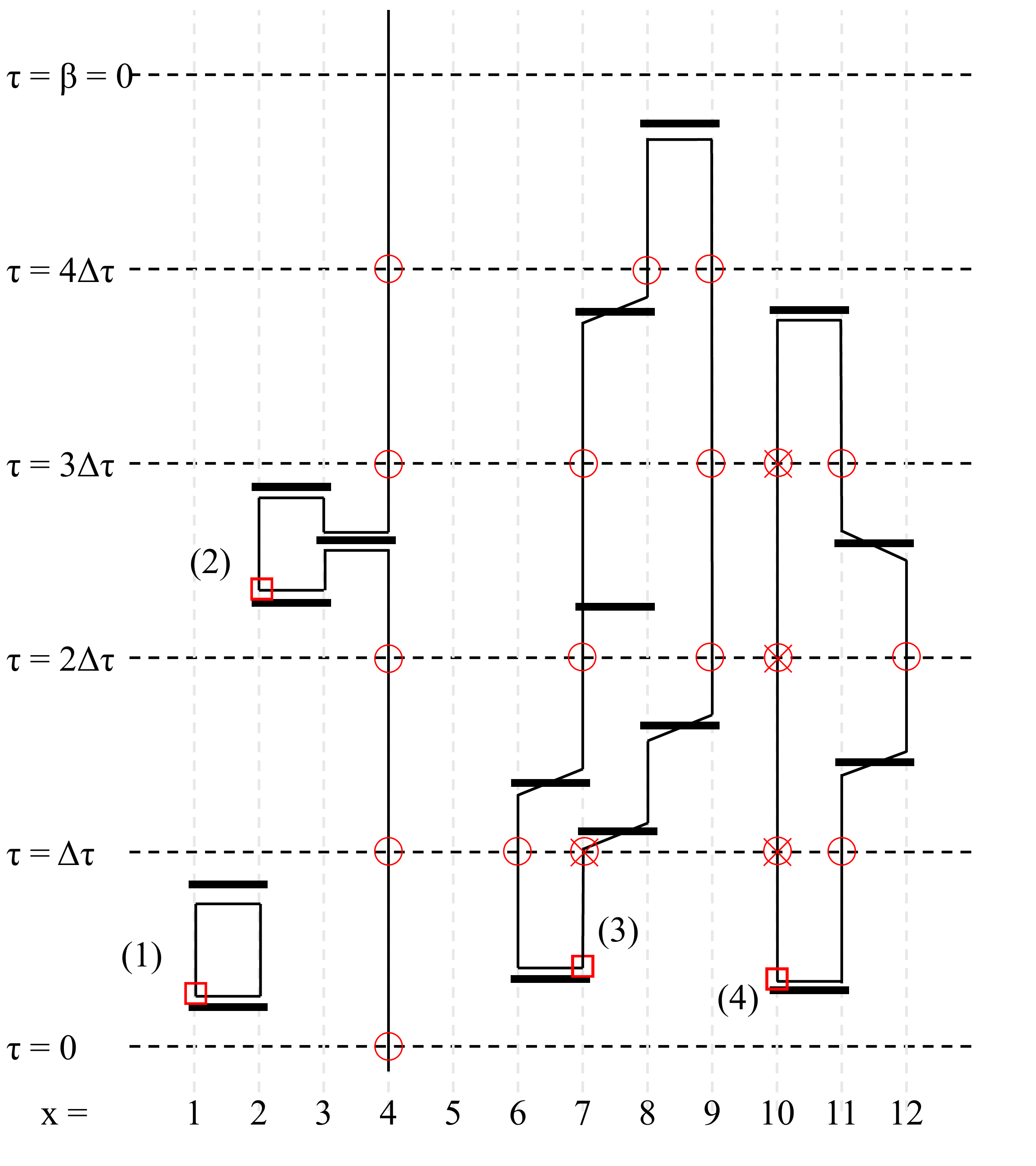}
  \caption{Example loops and their contributions to the correlator $G^\pm(r,\tau)$. $\rule{0.8em}{0.8ex}$ represent operators; solid lines are the loop paths; \textcolor{red}{$\square$} marks the initial vertex $v_0$; \textcolor{red}{$\times$} marks possible $\tau_0$ starting points, \textcolor{red}{$\bigcirc$} mark all contributing boundary points to the histogram in that loop. Loop \textbf{(1)} does not cross any time-slice boundaries, and so does not contribute to the histogram, although we still draw it as a normal MC update. Loop \textbf{(2)} does cross boundaries, but the initial leg chosen does not, and so it does not contribute to the histogram. Loop \textbf{(3)} starts on a link that crosses a single boundary at $x_0=7,\tau_0=\Delta\tau$. As we draw the loop, all boundary crossings (\textcolor{red}{$\bigcirc$}) contribute $G^\pm[\Delta r,\Delta\tau_k]$ $ +\!\!= n_{\text{legs}}$, where $n_{\text{legs}} = 4n_{H}$ for $n_{H}$ non-identity operators in the string. Loop \textbf{(4)} starts on a link that has multiple boundary crossings at $\tau_0=\Delta\tau,2\Delta\tau,3\Delta\tau$. One of these is randomly chosen as the starting location and $G^\pm[\Delta r,\Delta\tau_k]$ is incremented by $3n_{\text{legs}}$ for each boundary crossed by the loop to compensate for the initial choice.}
  \label{fig:CorrLoop}
\end{figure}

\begin{figure}[t] 
  \centering
  \includegraphics[width=0.8\columnwidth]{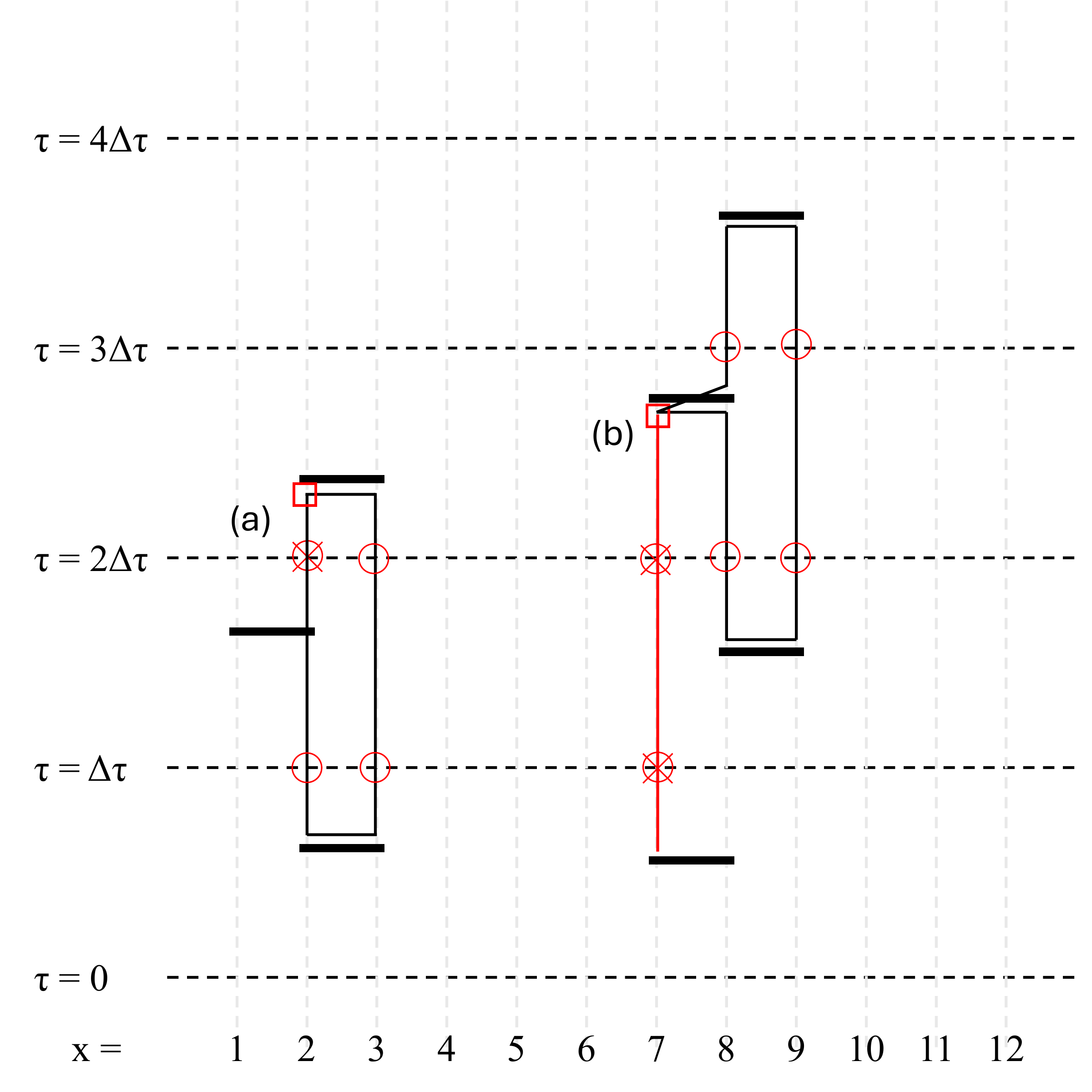}
  \caption{In the directed loops algorithm \cite{Syljuasen_PRE_2002}, there are two ways that loops can close. Loops begin with choosing a random vertex leg $v_0$ (\textcolor{red}{$\square$}) and moving to the exit leg for that vertex. A loop can close either reconnecting with the original vertex as (a) an entry leg or (b) an exit leg. Both loops shown above yield valid configurations in the closed space $\mathcal{C}$. For sampling in the extended configuration space $\mathcal{C'}$, only loop closings of type (a) are allowed, as type (b) causes biasing of the estimator. This is due to the method for defining a starting point $(x_0,\tau_0)$, which extends back along the leg from $v_0$ (red solid line). Allowing loop closures of type (b) leaves a loop that does not contain this leg, and so the construction is invalid. Loop closures of type (b) can be included if the tail (red solid line) is appended to the loop.}
  \label{fig:LoopClose}
\end{figure}

The above correlation function corresponds to inserting a spin-flip defect at some point in space-time $(0,0)$ and then ``healing'' it at a later point $(r,\tau_k)$, as shown in Fig.~\ref{fig:LoopDefect}. The propagation of these defects (a ``head'' and ``tail'' in worm algorithms) in SSE is handled naturally by directed-loop updates \cite{Dorneich_PRE_2001,Syljuasen_PRE_2002}. Unlike in the continuous-time formulations, we will only consider head/tail locations on the well-defined $\tau_k$ points.

We now derive the MC estimator for the correlation function above. Let $\mathcal{C}=\{\alpha,\mathcal{S}_{n_l}\}$ continue to denote the usual (closed) time-sliced SSE configurations with weights $W(\mathcal{C})$, so that $Z=\sum_{\mathcal{C}}W(\mathcal{C})$. We then define the expanded configuration space $\mathcal{C}' = \{\alpha, \mathcal{S}_{n_l}, \pm(r,\tau_k;r',\tau_{k'})\}$ where the new addition is the insertion of an $S^+$ and $S^-$ defect at locations $(r,\tau_k)$ and $(r',\tau_{k'})$ respectively. Note that $\mathcal{C}\subset\mathcal{C}'$, corresponding to the configurations where $r=r'$ and $\tau_k=\tau_{k'}$ (i.e. closed loops). During a directed-loop update, intermediate (open-loop) states correspond naturally to configurations in the extended space $\mathcal{C}'$ in which the head and tail defects are separated in space and imaginary time.

The weights in the extended configuration space are 
\begin{equation}
\begin{split}
W&(\mathcal{C}') 
    = w(\mathcal{C}) 
    \left\langle \alpha \left|
        \left(\prod_{l \ge k} \mathcal{S}_l \right)
        S^{+}(r)
        \left(\prod_{l < k} \mathcal{S}_l \right)
        S^{-}(0)
    \right| \alpha \right\rangle .
\end{split}
\end{equation}
This is precisely the numerator of the correlation function which we seek to measure. 
We can then identify the MC estimator as
\begin{equation}
f(\mathcal{C}'; r, \tau_k) =
   \begin{cases}
   1, & \parbox[t]{0.45\linewidth}{Open loop with head/tail separated by $(r, \tau_k)$,} \\[15pt]
   0, & \text{All other loops.}
   \end{cases}
\end{equation}

\begin{figure*}[t]
  \centering
  \includegraphics[width=\textwidth]{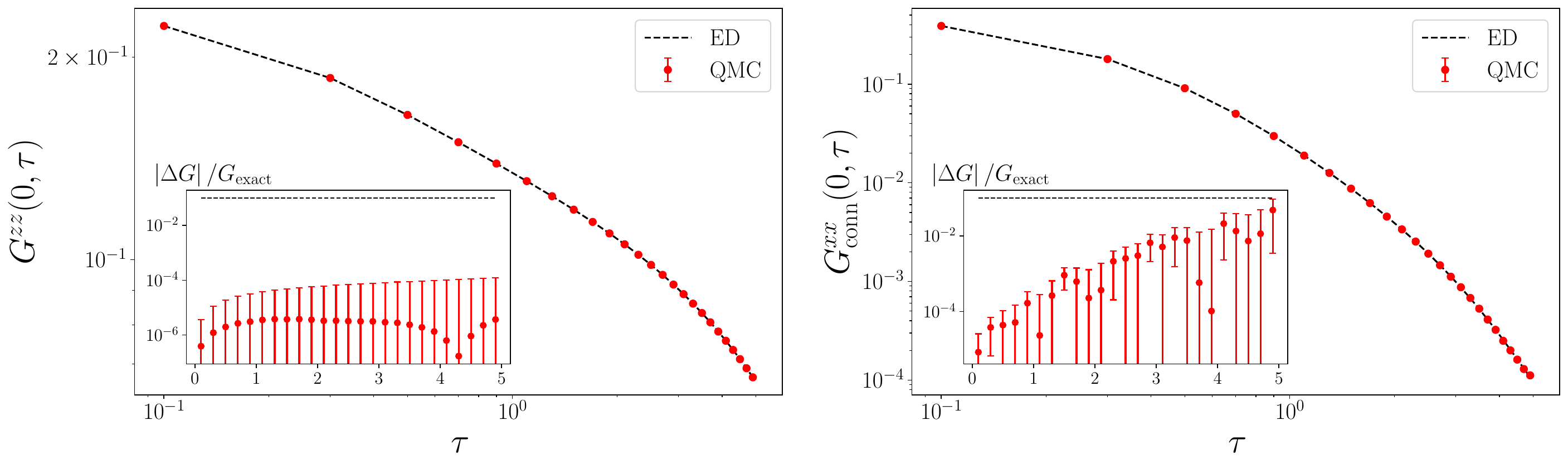}
  \caption{A comparison of the (left) same-site diagonal correlation function $G^{zz}(r=0,\tau)$ and (right) off-diagonal correlation function $G^{xx}_{\mathrm{conn}}(r=0,\tau)$ for the SSE QMC method detailed in Sec. \ref{OinH} and exact diagonalization (ED). Correlation functions are computed in the 1D TFIM for a small system $L=10$ at the critical point $h=1$ and inverse temperature $\beta = 4L$. Inset plots show relative error $|\Delta G|/G_{\text{exact}}$.}
  \label{fig:TFIM_ED}
\end{figure*}

Implicit in this definition is that the loop has non-zero weight in the extended partition function. $S^+/S^-$ must be inserted on $\downarrow/\uparrow$ spin states respectively. Contributions for $\bracket{S^+(r,\tau_k)S^-(0,0)}$ and $\bracket{S^-(r,\tau_k)S^+(0,0)}$ are accumulated separately based on their placement in the operator string, where $G^\pm(r,\tau_k) = G^\mp(r,\beta-\tau_k)$ are related by periodicity of the trace. 
We can therefore measure the off-diagonal correlation function by sampling configurations in the extended (open loop) space during the directed-loop update. For these configurations, we average over the estimator,
\begin{equation}
    \bracket{G_\pm(r, \tau_k)} \propto \bracket{f({\mathcal{C}';r, \tau_k})}_{W'},
\end{equation}
where $W'$ explicitly denotes that this is over open loop configurations. Due to the simplicity of the estimator, this has a very simple practical implementation. During the directed-loop update, we choose a random boundary $(r_0,\tau_{k,0}) \in (0\ldots N-1, 0\ldots m-1)$ to insert a pair of $S^+S^-$. We then draw the loop according to the normal rules derived from the directed loop equations. Unlike previous SSE estimators, this procedure does not require summation over the full operator string or explicit time-ordering of operators. Whenever the head passes over a new boundary $(r',\tau_{k'})$, we increment histograms

\begin{subequations}\label{Eq:BoundaryStart}
\begin{align}
    C_\pm[\Delta r, \Delta\tau_k] &= C_\pm[\Delta r, \Delta\tau_k] + 1\\
    C_\mp[\Delta r, \beta-\Delta\tau_k] &= C_\mp[\Delta r, \beta-\Delta\tau_k] + 1,
    \end{align}
\end{subequations}
for $\Delta r = r'-r_0$, $\Delta \tau_k = \tau_{k'}-\tau_{k,0}$ being the correct periodic separations between $S^-$ and $S^+$. These histograms are then exactly the unnormalized Green's functions we want. 

\subsection{Importance Sampling Improved Estimator}

In practice, the SSE algorithm is not well suited to picking a random boundary to start at. Instead, we usually begin drawing a loop at a vertex leg $l_0$, which belongs to a link (between two vertex legs of different vertices) that may or may not cross a time-slice boundary. Choosing a boundary to start at would require a search of what leg the boundary lies on. While this is not prohibitively expensive, we here present an importance sampling algorithm which can be more efficient than the more general case described above. A visualization of this scheme and possible loops is shown in Fig.~\ref{fig:CorrLoop}. 

(1) Choose a random leg $l_0$ from the vertex graph from which we begin our loop. Note this is done with probability $1/n_{\text{legs}}(\mathcal{C})$, where $n_{\text{legs}}(\mathcal{C}) \propto n_{H}$ and $n_H$ is the number of non-identity operators in the string for this configuration $\mathcal{C}$.

(2) If the link from $l_0$ does not cross any time-slice boundaries, then this loop will not contribute to the histogram, \textit{even if other legs in this loop cross boundaries}; See loops \textbf{(1)} and \textbf{(2)} in Fig.~\ref{fig:CorrLoop}.  If $l_0$ does cross some number of time-slice boundaries $n_x(l_0) \neq 0$, then uniformly choose one of those boundaries to be the location for the ``imagined'' insertion (for the purpose of the measurement) with probability $1/n_x$.

(3) Traverse the loop (even if it will not contribute to the histogram). If $n_x(l_0) \geq 1$, then every time we cross another time-slice boundary with separation $(r, \tau_k)$ from the starting point, increment the correct histogram determined by the head/tail of the loop by $n_x(l_0)n_{\text{legs}}(\mathcal{C})$. In practice, increment both of
\begin{subequations}
\begin{align}
C_\pm(r, \tau_k) = & C_\pm( r, \tau_k) + n_x(l_0)n_{\text{legs}}(\mathcal{C}) \\
C_\mp(r, \beta-\tau_k) = & C_\mp(r, \beta-\tau_k) + n_x(l_0)n_{\text{legs}}(\mathcal{C}).
\end{align}
\end{subequations}
Because the $1/n_x(l_0)n_{\text{legs}}(\mathcal{C})$ probability of choosing the leg $l_0$ and boundary $(r_0,\tau_{k_0})$ as a starting point is configuration dependent, it must be compensated within the loop. This is achieved through incrementing by factor $n_x(l_0)n_{\text{legs}}(\mathcal{C})$. In practice, $n_{\text{legs}}(\mathcal{C})$ is a large number (of order $\beta N$), but can be controlled by normalizing by a suitable configuration independent constant like $4mM$, the total number of legs including identities. Then the increment is by $n_x(l_0)n_{\text{legs}}(\mathcal{C}) / 4mM$, which is an $\mathcal{O}(1)$ number. 

\begin{figure*}[t]
  \centering
  \includegraphics[width=\textwidth]{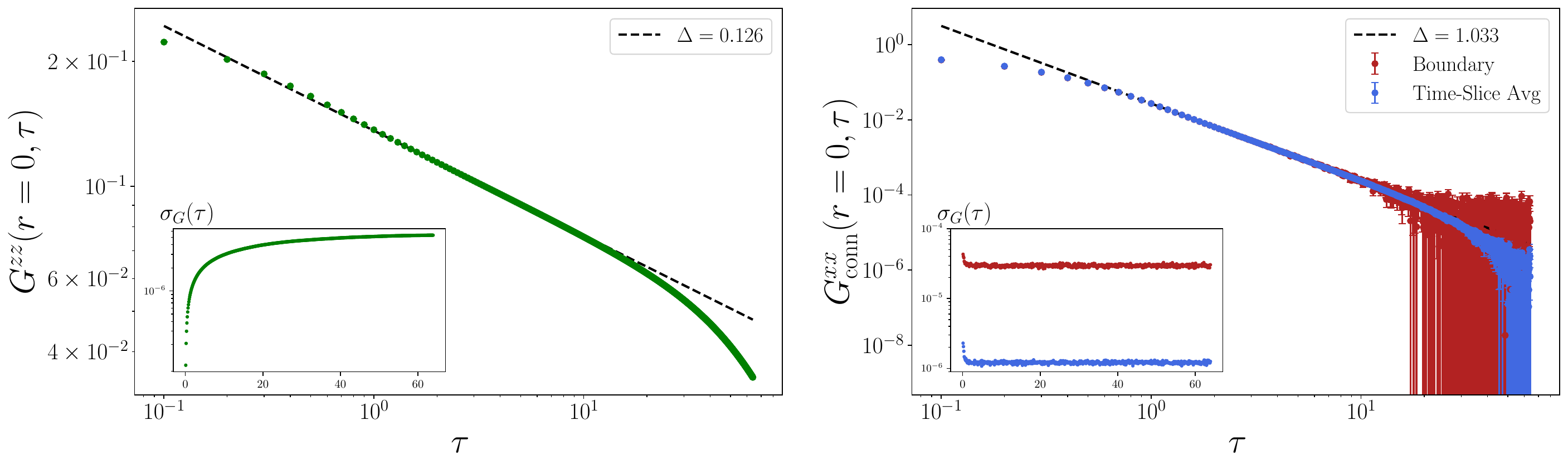}
  \caption{Same-site imaginary-time correlation functions for the $L=128$ site TFIM chain for both (left) diagonal $G^{zz}(\tau)$ and (right) off-diagonal $G^{xx}_{\text{conn}}(\tau)$ operators at $\beta=8L$ and $\Delta\tau = 0.1$. Scaling dimensions are extracted via a power-law fits to the curves. The two off-diagonal estimators defined in Sec.~\ref{OinH} are both shown in the right panel, with an order-of-magnitude reduction in the error bar for the time-slice averaged estimator. }
  \label{fig:TFIM_128}
\end{figure*}

(4) Repeat this process for all loops drawn in the update step. 

There is one more subtlety to note in the construction of loops. In the directed loops algorithm \cite{Syljuasen_PRE_2002}, loops can close by reconnecting with the original vertex leg either as an entry or exit leg. See Fig.~\ref{fig:LoopClose} for an example of each type of closing. Both methods are valid for generating new configurations in the ``closed'' configuration space $\mathcal{C}$. For sampling in the extended configuration space $\mathcal{C}'$, loop closures through an exit leg [type (b) in Fig.~\ref{fig:LoopClose}] must be treated carefully, as they lead to loops that do not contain the initial space-time site, rendering the estimator invalid. To include such closures, the ``tail'' (red solid line) must be included in the loop. These closures can also be excluded entirely with no bias introduced (simply continue to extend the loop until a closure of type (a) is encountered).

In both the boundary-start and importance-sampling methods, the histogram accumulated during loop propagation is unnormalized, and explicitly calculating the normalization is nontrivial due to loops of variable length. In practice, the normalization is often not important (only the rate of the decay is). In cases where a normalization is desired, one can fix the overall normalization afterwards by pinning a known value, e.g. $G_\pm(0,0)={1}/{2}$ for spin-$1/2$ with no external field, which is the method we use in the results section below.

Which method (boundary-start or importance sampling) is more efficient will depend on the model and parameters. While the importance-sampling method is expected to be more efficient near symmetry points (for example, near the isotropic Heisenberg point for the XXZ model), the boundary-start method is expected to be more efficient in regimes where the frequency of ``bounces'' in the directed-loop update (e.g. with a large field $h$ in XXZ) is large. 

\section{Results}\label{Results}

We first present results for small systems that can be checked against exact diagonalization and then consider examples of larger-scale calculations to demonstrate the precision that can be achieved with modest computing resources. 

\subsection{Transverse-Field Ising Model}

We first demonstrate the measurement of off-diagonal correlation functions for operators
$\mathcal{O}\in H$ using the one-dimensional transverse-field Ising model (TFIM). We compute the imaginary-time–resolved connected correlator
\begin{equation}
G^{xx}_{\text{conn}}(r,\tau)
    = \bracket{\sigma^x(r,\tau)\sigma^x(0,0)} - \bracket{\sigma^x}^2 ,
\end{equation}
as well as the diagonal correlation function
\begin{equation}
G^{zz}(r,\tau)
    = \bracket{\sigma^z(r,\tau)\sigma^z(0,0)}.
\end{equation}
$\sigma^z$ is diagonal, and therefore measured trivially using the method outlined in Sec.~\ref{DCorr}. Since $\sigma^x$ appears explicitly in the Hamiltonian, the $G^{xx}$ case tests the time-sliced SSE estimator for off-diagonal operators that are contained in $H$, as detailed in Sec. \ref{OinH}. The ferromagnetic TFIM is given by Eq.~(\ref{Eq:TFIM}). We focus on the critical point $h=1$, where correlations are long-ranged in both space and imaginary time. 

We compute both $G^{zz}$ and $G^{xx}_{\text{conn}}$ using the algorithms described above and compare the results to exact diagonalization (ED) for a small system of size $L=10$. In the time-sliced formulation, the off-diagonal estimator reduces to counting $\sigma^x$ operators adjacent to the time-slice boundaries $\tau_k$, allowing direct access to $\tau$-resolved correlators without additional summations over the operator string. We simulate at an inverse temperature $\beta = 4L$, and cut into $m=100$ time slices, which corresponds to $\Delta\tau = 0.4$. 

\begin{figure*}[t] 
  \centering
  \includegraphics[width=\textwidth]{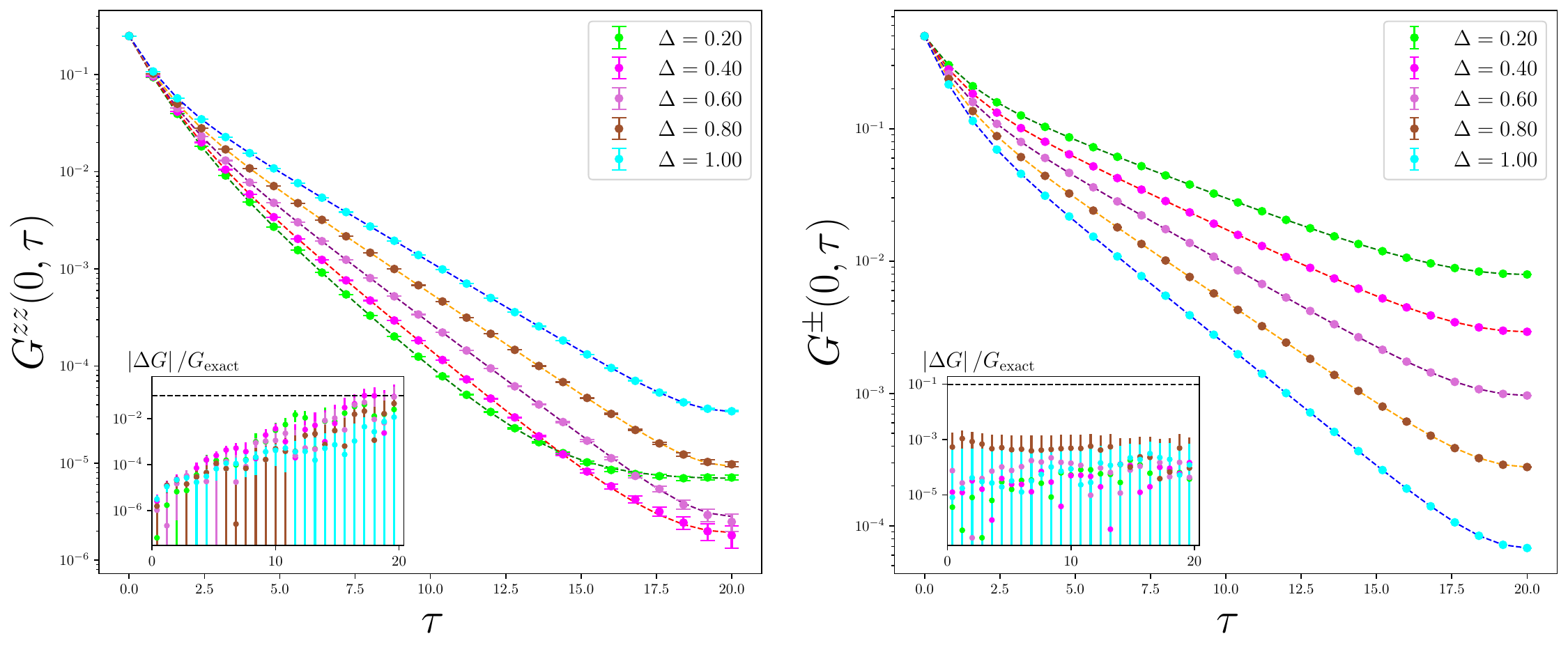}
  \caption{A comparison of same-site correlation functions $G^{zz}(r=0,\tau) = \bracket{S^z(0,\tau)S^z(0,0)}$ and $G^{\pm}(r=0,\tau) = \bracket{S^+(0,\tau)S^-(0,0)}$ for the SSE QMC method detailed in this work and exact diagonalization (ED). Correlation functions are computed in the 1D XXZ model with no external field for a small system $L= 10$ at inverse temperature $\beta = 4L$. Inset plots show relative error $|\Delta G|/G_{\text{exact}}$. }
  \label{fig:XXZ_ED}
\end{figure*}

The results are shown in Fig.~\ref{fig:TFIM_ED}. Same-site imaginary-time correlations $G(r=0,\tau)$ show excellent agreement with ED across the full range of separations, validating the correctness of the time-sliced SSE estimator for off-diagonal operators contained in the Hamiltonian. The relative error, $|\Delta G|/G_{\text{exact}}$ stays very small for $G^{zz}$ for all $\tau$ separations. For $G^{xx}_{\text{conn}}$, the relative error increases with $\tau$, and thus in order to use this data as input to SAC, we would cut off the correlation function at some $\tau_c < \beta/2$. A relative error of roughly $10\%$ is typically used as the cutoff for SAC, which we've highlighted with the dashed black line in the inset. 

We next look at the larger $L=128$ site TFIM chain, where we can extract the scaling dimensions from the imaginary time correlation functions. In Fig.~\ref{fig:TFIM_128}, we show the results for both the diagonal and off-diagonal correlation functions. We take $\beta = 8L$ and $\Delta\tau=0.1$. In both cases, we are able to extract scaling dimensions of $\Delta_\sigma=0.126$ and $\Delta_\epsilon = 1.033$ in good agreement with expected thermodynamic values of $\Delta_\sigma=1/8$ and $\Delta_\epsilon = 1$. At very large imaginary-time separations, the power-law decay of the correlations crosses over to exponential decay, as expected for a finite-size system. This is seen most clearly in the diagonal correlation function. Additionally, we show the benefits of the time-slice averaging procedure. There is at least an order of magnitude reduction in the standard deviation as compared to the boundary-valued estimator, and the regime in which the error bars are ``small'' extends to larger imaginary-time values. For extracting the decay (the scaling dimension), the small bias expected for the time-slice averaged estimator appears to make no difference. For each of the estimators, these results were obtained with roughly 1000 CPU hours. 

\subsection{1D XXZ Chain}

The XXZ model is an anisotropic Heisenberg model with Hamiltonian given by Eq.~(\ref{Eq:XXZ}). For $\abs{\Delta} > 1$ there is an (anti)ferromagnetic state depending on the value of J, while there is a gapless Luttinger liquid phase for $-1 < \Delta \leq 1$. $\Delta = 1, h=0$ recovers the isotropic Heisenberg model. At this point, the correlation functions
\begin{equation}
    \begin{aligned}
    G^{zz}(r, \tau) &= \bracket{S^z(r,\tau)S^z(0,0)}\\
        G^{\pm}(r, \tau) &= \bracket{S^+(r,\tau)S^-(0,0)}
    \end{aligned}
\end{equation}
are related simply by $2G^{zz}(r, \tau) = G^{\pm}(r, \tau)$. While we first benchmark against exact diagonalization for small chains, this isotropic point serves as a good consistency check for larger lattice sizes. 

The SSE implementation of the XXZ model is straightforward, and we follow the algorithm using directed loops from Ref. \cite{Syljuasen_PRE_2002}. The relevant matrix elements for the local operators acting on two sites are
\begin{equation}
\begin{gathered}
    \expec{\uparrow\downarrow}{H_b}{\downarrow\uparrow} = \expec{\downarrow\uparrow}{H_b}{\uparrow\downarrow} = \frac{1}{2}\\
    \expec{\uparrow\downarrow}{H_b}{\uparrow\downarrow} = \expec{\downarrow\uparrow}{H_b}{\downarrow\uparrow} = \frac{\Delta}{2}+\mathbf{\epsilon}+h_b\\
    \expec{\downarrow\downarrow}{H_b}{\downarrow\downarrow} = \mathbf{\epsilon}\\
    \expec{\uparrow\uparrow}{H_b}{\uparrow\uparrow} = \mathbf{\epsilon}+2h_b\\
\end{gathered}
\end{equation}
where $h_b = h/2dJ$ with $d$ the dimension of the lattice. 

\begin{figure*}[t] 
  \centering
  \includegraphics[width=\textwidth]{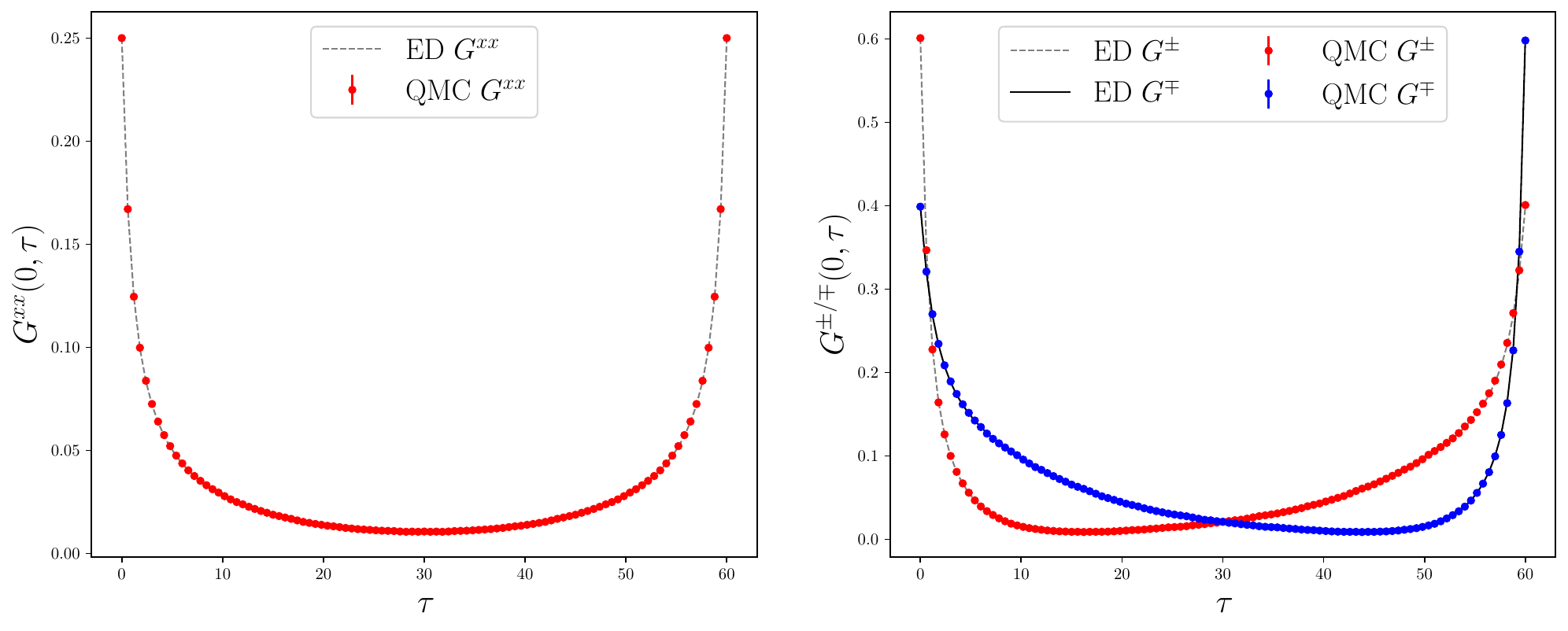}
  \caption{Imaginary-time same-site correlation functions (left) $G^{xx}(r=0,\tau)$ and (right) $G^{\pm/\mp}(r=0,\tau)$ for an XXZ spin chain of size $L = 10$ with an external field $h=0.5$. Simulations are at inverse temperature $\beta = 6L$, with $\Delta\tau = 0.6$. SSE data show excellent agreement with exact diagonalization. The transverse correlation function can be decomposed $G^{xx}(r,\tau) = [G^{\pm}(r,\tau) + G^{\mp}(r,\tau)]/4$, which are shown in the right plot. With an external field, the relation $G^{\pm}(r,\tau) = G^{\mp}(r,\beta-\tau)$ holds.}
  \label{fig:XXZh_ED}
\end{figure*}

First, we work only within the ``zero bounce'' regime by restricting ourselves to zero external field and $0 \leq \Delta \leq 1$. This is also the regime where $\bracket{S^z} = 0$, and so there is no distinction between $G_\pm$ and $G_\mp$. We take the minimum value for $\epsilon$, $\epsilon_{\text{min}} = {(1-\Delta)}/{4}$, in order to eliminate bounce moves. We have set $J=1$ as the energy scale. The Hamiltonian is then
\begin{equation}
    H = \sum_i\left[\frac{1}{2}\left(S^+_iS^-_{i+1} + S^-_iS^+_{i+1}\right) + \Delta\left(S_i^zS_{i+1}^z-\frac{1}{4}\right)+\epsilon\right]
\end{equation}

We follow the algorithms detailed in Sec.~\ref{DCorr} and Sec.~\ref{OnotinH} to measure the diagonal and off-diagonal correlation functions. Specifically, we here implement the importance-sampling method where loops are started on a chosen leg, rather than a boundary. We then compare to ED calculations, and the results are shown for a chain of length $L=10$ at different values of $\Delta$ in Fig.~\ref{fig:XXZ_ED}. We simulate at inverse temperature $\beta = 4L$, and with time slices $\Delta\tau = 0.4$ (ticks show every other point in Fig.~\ref{fig:XXZ_ED}). Note that we show $G^{\pm} = G^{\mp}$ here; both are equivalent as there is no external field. We find excellent agreement with ED for both $G^{zz}(\tau)$ and $G^{\pm}(\tau)$. As with the off-diagonal estimator for the TFIM in the previous section, the relative error increases with imaginary time $\tau$ for $G^{zz}(\tau)$, and we again plot the expected cutoff for SAC at $10\%$ relative error. The relative error for the off-diagonal correlation function is roughly flat at $|\Delta G|/G_{\text{exact}} \lesssim 0.1\%$ out to cutoff $\beta/2$. 

Next, we add an external field to the system. We simulate the chain at $\Delta=1$, $h=0.5$, and $\epsilon = 0$. This introduces bounce moves to the directed-loop update, and also breaks the symmetry such that now $G_\pm(r,\tau) = G_\mp(r,\beta-\tau) \not\propto G^{xx}$. In Fig.~\ref{fig:XXZh_ED} we show that we still have excellent agreement with ED in this most general case. The transverse correlator $G^{xx}(r,\tau)$ is obtained by accumulating all boundary crossings without distinguishing between the $S^+$ and $S^-$ ends of the loop, which is equivalent to averaging $\left(G^{\pm}(r,\tau) + G^{\mp}(r,\tau)\right)/4$. The final result for these correlators also matches ED, and shows the expected $G^{\pm}(r,\tau) = G^{\mp}(r,\beta-\tau)$ symmetry. 

\begin{figure*}[t] 
  \centering
  \includegraphics[width=\textwidth]{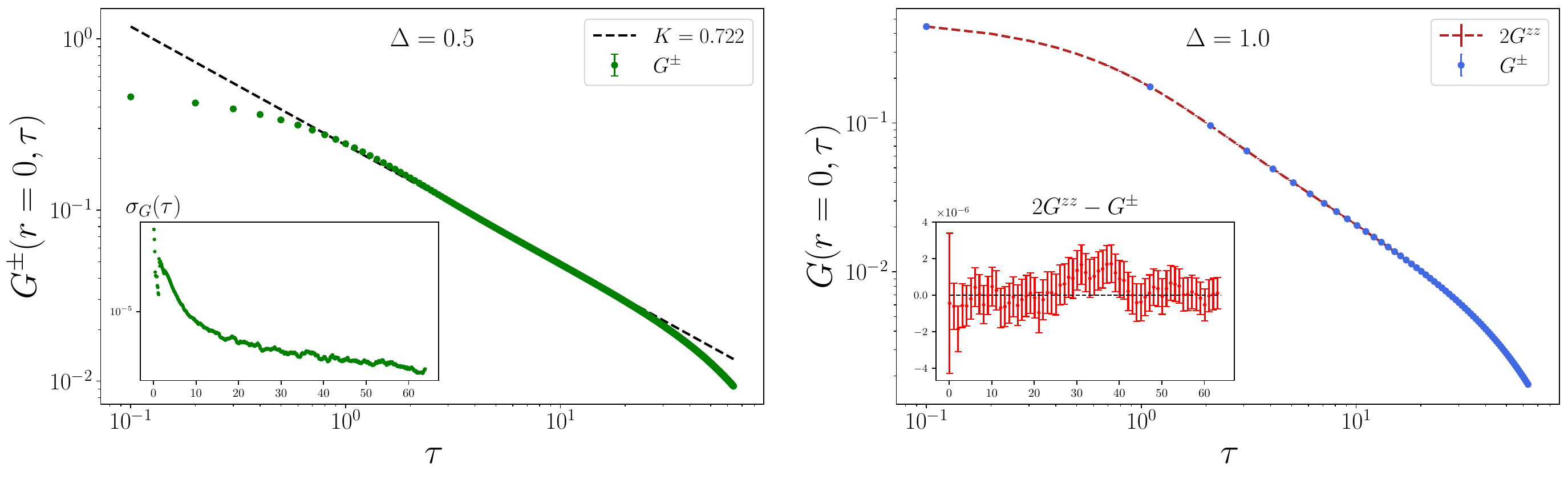}
  \caption{Same-site imaginary-time correlation functions for the $L=128$ site XXZ chain with no external field. (left) $G^\pm(\tau)$ for $\Delta = 1/2$, in the critical Luttinger-liquid phase. The Luttinger parameter $K = 0.722$ is extracted from the power-law decay. (right) $G^{zz}(\tau)$ and $G^\pm(\tau)$ for $\Delta = 1$, the isotropic point. The diagonal and transverse correlations match within statistical errors, as expected at the SU($2$) symmetric point.}
  \label{fig:XXZ_128}
\end{figure*}

We again look at the larger $L=128$ site chain to demonstrate the efficiency of the method. In Fig.~\ref{fig:XXZ_128}, we show the results for the XXZ chain in no external field for both $\Delta = 0.5$ and $\Delta = 1$, at $\beta = 8L$ and with $\Delta\tau = 0.1$. For $\Delta = 0.5$, the system is in the critical Luttinger-liquid phase and we expect power-law decay of the correlation functions. The Luttinger parameter $K$ can be extracted from the power-law decay, which for long imaginary-time separations goes like~\cite{Giamarchi}
\begin{equation}
    \bracket{S^+(\tau)S^-(0)} \propto \tau^{-{1}/{2K}}.
\end{equation}
We extract a value $K=0.722$, about $4\%$ off from the Bethe ansatz value of $3/4$. For $\Delta = 1$, we show how the method can be tested for consistency at system sizes larger than exact diagonalization can reach. At the isotropic point, the diagonal and transverse correlation functions are related by $2G^{zz}(\tau) = G^{\pm}(\tau)$. The computed correlation functions at this point for the $L=128$ chain show agreement to within statistical error bars across the entire range $\tau = [0,\beta/2]$. The long-chain results were obtained with roughly $2000$ CPU hours. 

\section{Discussion}\label{Discussion}

We have introduced a method for measuring general imaginary-time correlation functions within stochastic series expansion quantum Monte Carlo simulations. While we have focused on spin systems, the approach applies broadly to any sign-problem-free model. The diagonal and $\mathcal{O}\in H$ estimators rely on standard SSE machinery, while the $\mathcal{O} \notin H$ estimator additionally requires a directed-loop or worm-type update scheme. The central idea is to combine an explicit imaginary-time slicing of the SSE operator string with an interpretation of directed-loop updates as sampling an extended configuration space associated with off-diagonal correlation functions. Although each of these ingredients has appeared previously in different contexts, their combination yields a new and efficient scheme for measuring imaginary-time correlations with explicit $\tau$ resolution.

The resulting algorithms are computationally efficient. For diagonal operators and off-diagonal operators appearing in the Hamiltonian, the computational cost scales as $\mathcal{O}(m L^d\log m L^d)$ for $m=\beta/\Delta_\tau$ (with $\Delta_\tau$ the desired imaginary-time resolution) due to the use of fast Fourier transforms, while for off-diagonal operators not contained in the Hamiltonian the cost scales as $\mathcal{O}(m L^d)$. The usual SSE update scheme also has $\mathcal{O}(m L^d)$ scaling, and so the measurement of correlation functions introduces only a minimal computational overhead in this case. This is a notable improvement over earlier SSE-based approaches, which typically required explicit summation over the full operator string to evaluate imaginary-time correlation functions.

Recently, complementary approaches to measuring off-diagonal observables in QMC have been proposed, including a reweighting and annealing method \cite{Wang_Nature_2026} and a generalized reduced-density matrix framework \cite{Wang_arxiv_2026}. These methods offer broad generality and provide powerful routes to observables that may be difficult to access directly, formulating expectation values and correlation functions as ratios of partition functions or by replacing the partition function itself as the sampled object. The approach presented here differs in scope and emphasis. By exploiting the structure of SSE and directed-loop updates, we construct a direct estimator for two-point correlation functions that can be accumulated within a single simulation ensemble, without the need for annealing, reweighting, or modified sampling targets. While the above approaches offer greater generality, the present method is optimized for two-point functions and introduces only minimal computational overhead beyond standard loop updates, avoiding additional sources of statistical uncertainty.

Closely related ideas have also appeared in recent work on single-hole spectral functions in $t$-$J$-type models~\cite{Yang_2025a,Yang_2025b}, where a hole is injected into a time-sliced SSE configuration and its imaginary-time evolution is sampled. That approach targets fermionic single-particle Green’s functions and is complementary to the present work, which focuses on two-point spin and bosonic correlators accumulated within directed-loop updates. 

The efficient access to full imaginary-time correlation functions makes this method particularly well suited for subsequent analytic continuation procedures, for example using stochastic analytic continuation (SAC), aimed at extracting real-frequency dynamical response functions. In previous SSE-based studies, such analyses were largely restricted to $SU(2)$ symmetric models or to diagonal structure factors such as $S^z(k,\omega)$. The present method removes these restrictions and enables the study of anisotropies, external fields, and transverse dynamical correlations within the same computational framework.

\begin{acknowledgments}
This work was supported by the Simons Foundation under Grant No.~511064 (A.W.S. and R.F.), a Fellowship from the John Simon Guggenheim Memorial Foundation (A.W.S.), NSTC of Taiwan Grant No.~113-2112-M-002-033-MY3
(Y.J.K.), and Singapore Ministry of Education (MOE) Academic Research Fund Tier 3 Grant No.~MOI-MOET32023-0003
(S.Y.). Computing resources were provided by Boston University’s Research Computing Services. NTU's High Performance Computing Centre
computing also provided resources, facilities, and services that have contributed to this work.  
\end{acknowledgments}

\bibliography{references.bib}

\end{document}